\documentclass{LMCS}

\def\doi{7 (4:01) 2011}
\lmcsheading%
{\doi}
{1--25}
{}
{}
{Oct.~16, 2010}
{Oct.~26, 2011}
{}

\usepackage{enumerate,hyperref}
\theoremstyle{plain}
\theoremstyle{definition}

\newtheorem{note}[thm]{Note}

\def\uun{\rlap{$\uparrow$}\raise.4ex\hbox{$\uparrow$}}
\def\ddn{\rlap{$\downarrow$}\raise.4ex\hbox{$\downarrow$}}
\def\Uun{\raise.4ex\hbox{$\Uparrow$}}

\begin{document}

\title[Computational Models of Certain Hyperspaces  of Quasi-metric Spaces]
{Computational Models of Certain Hyperspaces of Quasi-metric Spaces}

\author[M.~Ali-Akbari]{Mahdi Ali-Akbari}	
\address{Department of Mathematics, Semnan University, P.O. Box 35195-363, Semnan, Iran and
School of Mathematics, Institute for Research in Fundamental Sciences (IPM), Tehran, Iran.}	
\email{m\_aliakbari@aut.ac.ir}  
\thanks{The authors are partially supported by IPM, grants No. 89540064 and No. 89030120.}	

\author[M.~Pourmahdian]{Massoud Pourmahdian}	
\address{School of Mathematics and Computer Science, Amirkabir University of Technology, Tehran, Iran  and 
School of Mathematics, Institute for Research in Fundamental Sciences (IPM), Tehran, Iran.}	
\email{pourmahd@ipm.ir}  



\begin{abstract}

 \noindent In this paper, for a given sequentially Yoneda-complete $T_1$
quasi-metric space $(X,d)$, the domain theoretic models of the
hyperspace ${\mathcal K}_0(X)$ of nonempty compact subsets of $(X,d)$ are
studied. To this end, the $\omega$-Plotkin domain of the space of
formal balls ${\bf B}X$, denoted by ${\bf C}{\bf B}X$ is
considered. This domain is given  as the chain completion of the
set of all finite subsets of ${\bf B}X$ with respect to the
Egli-Milner relation. Further, a map $\phi:{\mathcal
K}_0(X)\rightarrow{\bf C}{\bf B}X$ is established and proved that
it is an embedding whenever ${\mathcal K}_0(X)$ is equipped with the
Vietoris topology and respectively ${\bf C}{\bf B}X$ with the
Scott topology. Moreover, if any compact subset of $(X,d)$ is
$d^{-1}$-precompact, $\phi$ is an embedding with respect to the
topology of Hausdorff quasi-metric $H_d$ on ${\mathcal K}_0(X)$.
Therefore, it is concluded that $({\bf C}{\bf
B}X,\sqsubseteq,\phi)$ is an $\omega$-computational model for the
hyperspace ${\mathcal K}_0(X)$ endowed with the Vietoris and
respectively the Hausdorff topology.

Next, an algebraic sequentially Yoneda-complete quasi-metric $D$
on ${\bf C}{\bf B}X$ is introduced in such a way that the
specialization order $\sqsubseteq_D$ is equivalent to the usual
partial order of ${\bf C}{\bf B}X$ and, furthermore, $\phi:({\mathcal
K}_0(X),H_d)\rightarrow({\bf C}{\bf B}X,D)$ is an isometry. This
shows that $({\bf C}{\bf B}X,\sqsubseteq,\phi,D)$ is a
quantitative $\omega$-computational model for $({\mathcal
K}_0(X),H_d)$.

\end{abstract}

\maketitle

\section*{Introduction}

In this paper, we further continue a project carried out to
investigate connections between domain theory and quasi-metric
spaces \cite{1}. Here, we provide some domain theoretic
(computational) models for the hyperspace of nonempty compact
subsets of quasi-metric spaces.  On one hand, the recent
applications of quasi-metric spaces in different subjects of
computer science, e.g. denotational semantics of programming
languages, complexity and dual-complexity spaces and complexity
distances between algorithms (\cite{rom-val, rod-sch-val,
gar-rom-sch, rom-san-val, rom-sch, rodri-rom-val }) and, on the
other hand, a new insight of the domain theoretic point of view
into the theory of hyperspaces and its new applications within
mathematics, e.g. discrete dynamical systems, measure and
integration theory (\cite{edalat1,edalat2,edalat3}), motivate
establishing computational models of these structures.

Finding a domain theoretic (computational) model for a topological
space $(X,\tau)$ amounts to providing a suitable partially ordered
set $(P,\sqsubseteq)$ together with a topological embedding $\phi$
from $(X,\tau)$ to $(P,\sqsubseteq)$ endowed with the Scott
topology, denoted by $\sigma$. This is a variant of a fundamental
problem in domain theory, called the maximal point space problem,
which demands a homeomorphism between $(X,\tau)$ and the space of
maximal point of $(P,\sqsubseteq)$. The study of computational
models for various type of topological spaces goes back to the
works of Edalat  and respectively Blanck \cite{edalat1, Blanck,
Blanck1}. Later, the maximal point space problem was explicitly
formulated and became a subject of intensive investigations by
many authors \cite{Law, Lutzer, Martin, Rutten}. Some special
cases of this problem have satisfactory solutions
\cite{2,kop-kunz}.

The domain theoretic construction $\mathbf{B}X$ of the space of
formal balls, introduced by Edalat and Heckmann, provides a concrete
(computational) model for a metric space $(X,d)$ \cite{edalat}. The
importance of this construction is that, first of all, it connects
some metric properties of $(X,d)$ to the order theoretic properties
of $\mathbf{B}X$. Secondly, it ties the above notion of
computational model to the notion of computability for metric space
$(X,d)$ \cite{edalat4,Laws}. The notion of formal balls is also
defined in the same way for a quasi-metric space $(X,d)$ and the
order theoretic properties of ${\bf B}X$ are tightly connected to
the topological properties of $(X,d)$ \cite{1,Rom-Val,Rom-Val1}. In
particular, for a $T_1$ quasi-metric space $(X,d)$ is sequentially
Yoneda-complete if and only if ${\bf B}X$ is a directed complete
partially ordered set.

Edalat and Heckmann also constructed the Plotkin powerdomain
${\mathcal P}{\bf B}X$ of the space of formal balls of a metric space
$(X,d)$ and showed that there is a one-to-one correspondence
between the nonempty compact subsets of $(X,d)$ and the maximal
elements of the Plotkin powerdomain of ${\mathbf{B}}X$. As an
application, a domain theoretic proof was given for a classical
result of Hutchinson (\cite{hutchinson}) which states that if
$(X,d)$ is complete, then any hyperbolic iterated function system
has a unique non-empty compact attractor. It can be shown that
this construction is a computational model for the hyperspace of
nonempty compact subsets of $X$, denoted by ${\mathcal K}_0(X)$, with
the Vietoris or equivalently the Hausdorff topology. This fact was
also proved in a different way by Martin in \cite{martin1}. His
interesting idea is based on the existence of a certain
measurement, called Lebesgue measurement, on any domain  $D$ which
models the metric space $(X,d)$. Subsequently, Liang and Kou in
\cite{L-K} generalized these results to continuous dcpo's which
have the Lawson condition, i.e. the Lawson  and Scott topologies
coincide on the space of maximal points. Indeed, under the Lawson
condition, it is proved that there is a homeomorphism between the
space of nonempty compact subsets of maximal points of a
continuous dcpo $D$ endowed with the Vietoris topology and the
space of maximal points of Plotkin powerdomain $D$ equipped with
the induced Scott topology. More recently, in another line of
research, Berger et al. (\cite{berger}) showed that for any $T_1$
topological space which is represented by an $\omega$-domain $D$,
the hyperspace of its nonempty compact subsets can be represented
by the Plotkin powerdomain ${\mathcal P}D$ of domain $D$. This result
was made possible by a theorem of Smyth (\cite{smyth}) which
states that for any $\omega$-continuous dcpo $D$ the space $({\mathcal
P}D,\sigma)$ is homeomorphic to the space of $D$-lenses, i.e.
nonempty compact subsets of domain $D$ which are intersection of
a closed set and a saturated set, endowed with the Vietoris
topology.

In the present work, we study computational models of the hyperspace
$\mathcal{K}_0(X)$ of a $T_1$ quasi-metric space $(X,d)$ equipped
with the Vietoris and respectively the Hausdorff topology,
proving that whenever $(X,d)$ satisfies certain completeness
properties, e.g. Yoneda and respectively Smyth  completeness,
this space has a computational model. It is worth mentioning that
the space of formal balls of a quasi-metric space does not
generally satisfy the Lawson nor countable based conditions.
Therefore, the results of Liang and Kou \cite{L-K} and Berger et
al. \cite{berger} do not apply to the present context. Also,
unlike the metric case, there is no natural candidate for a
measurement on the space of formal balls of a quasi-metric space
and hence the method used by Martin in \cite{martin1} cannot  be
applied here either.

Edalat and Heckmann used the Plotkin powerdomain
${\mathcal P}{\bf B}X$ given as  the ideal completion of the abstract
basis of finite subset of ${\mathbf{B}}X$, ${\mathcal P}_{fin}{\bf
B}X$, with respect to the Egli-milner relation, $\prec_{EM}$, to
present a computational model of $\mathcal{K}_0(X)$, for every
metric space $(X,d)$. To this end, they employed  the symmetry
axiom of metric $d$, to get a key fact that any maximal ideal has
a cofinal $\omega$-chain. In the case of quasi-metric spaces, the
Plotkin powerdomain ${\mathcal P}{\bf B}X$ can also be defined,
though, the lack of symmetry for the quasi-metric $(X,d)$
prevents us from finding cofinal  $\omega$-chains in maximal
ideals.  That is why we prefer to work directly with the
$\omega$-chains and this leads us to the chain-completion
construction instead.

So, for a $T_1$ quasi-metric space $(X,d)$, we consider the space
${\mathbf{B}}X$ of formal balls and let ${\mathbf{C}\mathbf{B}}X$
be the chain completion of $({\mathcal P}_{fin}{\bf B}X,\prec_{EM})$.
This construction is called the $\omega$-Plotkin domain. By the
general construction of chain completion,
${\mathbf{C}\mathbf{B}}X$ is a continuous $\omega$-dcpo, i.e. a
continuous poset in which every $\omega$-chain has a least upper
bound. Now, to achieve our purpose in finding a computational
model, we define a one-to-one map
$\phi:\mathcal{K}_0(X)\rightarrow {\mathbf{C}\mathbf{B}}X$, which
is an embedding if we consider the Vietoris topology on
$\mathcal{K}_0(X)$ and assume that $(X,d)$ is a sequentially
Yoneda-complete $T_1$ quasi-metric space. Moreover, $\phi$ is an
embedding with respect to the topology of the Hausdorff quasi-metric
$H_d$ on $\mathcal{K}_0(X)$ if any compact subset of $X$ is
$d^{-1}$-precompact. Therefore, $({\mathbf{C}\mathbf{B}}X,\phi)$
serves as an $\omega$-computational model for $\mathcal{K}_0(X)$
endowed with the mentioned topologies. Although it is not known
whether ${\mathbf{C}\mathbf{B}}X$ is a dcpo and therefore a
computational model of  $\mathcal{K}_0(X)$, nevertheless, thanks
to Fact \ref{idealcompletion} and Theorem \ref{dcpo},  the ideal
completion of ${\mathbf{C}\mathbf{B}}X$ gives a computational
model for  $\mathcal{K}_0(X)$.

In section 5, we take another well-known notion of computational
model, called the quantitative $\omega$-computational model. This
is an $\omega$-computational model $(P,\sqsubseteq,\phi)$ carrying
an additional quasi-metric $D$ such that $\phi$ is an isometry from
$(X,d)$ into $(P,D)$ together with some extra conditions which
capture the order structure of $(P,\sqsubseteq)$ (Definition
\ref{quantitative}). A modified version of this notion can be
found in \cite{Rom-Val,Rutten,Schellekens,Was}. We prove that in
fact ${\mathbf{C}\mathbf{B}}X$ is a quantitative
$\omega$-computational model for $(\mathcal{K}_0(X),H_d)$, by
constructing a quasi-metric $D$ on ${\mathbf{C}\mathbf{B}}X$. To
this end, we consider a quasi-metric $q$ defined by Romaguera and
Valero (\cite{Rom-Val}) on ${\bf B}X$. The primary reason to
choose this quasi-metric on $\mathbf{B}X$ is that
$(\mathbf{B}X,q)$ is a quantitative computational model for
$(X,d)$. Therefore, its specialization order $\sqsubseteq_q$ is
equivalent to the partial order of $\mathbf{B}X$. Consequently,
one could naturally extend $q$ to the Hausdorff quasi-metric
$H_q$ on ${\mathcal P}_{fin}{\bf B}X$ of the finite subsets of ${\bf
B}X$, whose main property is that it induces the Egli-Milner
relation on ${\mathcal P}_{fin}{\bf B}X$. Subsequently, the
quasi-metric $H_{q}$ can be lifted up to a quasi-metric $D$ on
${\mathbf{C}\mathbf{B}}X$ in such a way that the ordered
structures $({\mathbf{C}\mathbf{B}}X,\sqsubseteq_D)$ and
$({\mathbf{C}\mathbf{B}}X,\sqsubseteq)$ coincide. Once $D$ is
established one can show that $({\mathbf{C}\mathbf{B}}X,D)$ is a
Yoneda-complete space and in fact Yoneda-completion of $({\mathcal
P}_{fin}{\bf B}X, H_q)$. This makes $({\bf C}{\bf
B}X,\sqsubseteq,\phi,D)$ a quantitative $\omega$-computational
model for $({\mathcal K}_0(X),H_d)$.

We, finally, conclude this paper by comparing the Plotkin
powerdomain and the $\omega$-Plotkin domain constructions. We
prove that if $(X,d)$ is either, Smyth-complete and all of its
compact subsets are $d^{-1}$-precompact, or an $\omega$-algebraic
Yoneda-complete space, then  the Plotkin Powerdomain ${\mathcal
P}\mathbf{B}X$ is order-isomorphic to $\mathbf{C}\mathbf{B}X$.

\section{Preliminaries}

We assume the reader is familiar with the basic definitions and
facts about domain theory which can be found in
(\cite{abramsky,compendium}), though, we briefly explain some of
the definitions and facts which are more crucial in this note.

Let $(P,\sqsubseteq)$ be a partially ordered set (abbr. by poset).
The binary relation $\prec$ is called an {\em auxiliary} relation on the poset
$(P,\sqsubseteq)$ if (1) $p\prec p$ implies $p\sqsubseteq p$,
(2) $p\sqsubseteq s\prec r\sqsubseteq q$ implies $p\prec q$ and
(3) satisfies the interpolation property, i.e. for any finite subset
$M$ of $P$ and $p\in P$, if for every $m\in M$, $m\prec p$ then there
exists some $q\in P$ such that $m\prec q\prec p$, for every $m\in M$.
The pair $(P,\prec)$ is called an {\em abstract basis}, if $\prec$ is
a transitive relation which also satisfies the interpolation
property. A nonempty directed lower subset $I$ of $P$ is called a
{\em round ideal} if for any $x\in I$ there is $y\in I$ such that
$x\prec y$. The set of all round ideals of $P$ partially ordered
by $\subseteq$ is called the {\em ideal completion} of $P$, denoted by
$Idl(P)$. Let $\uun p=\{q:\ p\prec q\}$ and $\ddn p =\{q:\
q\prec p\}$. An auxiliary relation is called {\em approximating} if
$\ddn p \subseteq \ddn q$ implies $p\sqsubseteq q$.
One can see that the set $\{\uun p:\ p\in P\}$ forms a basis for a
topology called the {\em pseudoScott topology} on $P$, denoted by
$\mathbf{P}\sigma$. The following fact is needed for the proof of
Theorem \ref{domrep}.

\begin{fact}\label{idealcompletion}
Let  $(P,\sqsubseteq)$ be a poset with an auxiliary relation $\prec$. Then
\begin{enumerate}[(1)]
\item $(Idl(P),\subseteq)$ is a continuous dcpo.

\item If $\prec$ is approximating on $P$, then the map
$j: P \rightarrow Idl(P)$ defined by $j(p)=\ddn p $
is an embedding of $(P,\mathbf{P}\sigma)$ into $(Idl(P),\sigma)$
where $\sigma$ denotes the Scott topology.

\item If $\prec$ is approximating and all $\prec$-directed sets
of $P$ have upper bounds, then $j(\max{P})=\max{Idl(P)}$.
\end{enumerate}
\end{fact}
\proof See \cite{kop-kunz}, Theorem 2.3.\qed

Below, we fix the key notion of a computational model for a $T_0$
topological space. Before that, recall any $T_0$ topology $\tau$
on a space $X$ induces a partial order $\sqsubseteq_\tau$, called
the  specialization order, which is defined by
\[
x\sqsubseteq_\tau y \ \ \ \Leftrightarrow \ \ \ x\in cl_\tau {y},
\]
for all $x,y\in X$. $cl_\tau {y}$ stands for the closure of $y$ with
respect to $\tau$. Also, a partially ordered set $(P,\sqsubseteq)$
is an $\omega$-dcpo if every $\sqsubseteq$-ascending sequence has a least upper bound (see \cite{knijnenburg}).

\begin{defi}\label{computational}
A triple $(P,\sqsubseteq,\phi)$ is a {\em
($\omega$-)computational model} for $(X,\tau)$ whenever
\begin{enumerate}[(1)]
\item $(P,\sqsubseteq)$ is a continuous ($\omega$-)dcpo.
\item $\phi$ is a topological embedding from $(X,\tau)$ into $(P,\sqsubseteq)$ endowed with the Scott topology.
\item $\phi(Max(X,\sqsubseteq_\tau))= Max(P,\sqsubseteq)$.
\end{enumerate}
\end{defi}

Blanck in \cite{Blanck} considered this definition as a domain
representation for $(X,\tau)$ without mentioning  the third
condition. If we restrict ourselves to $T_1$ topological spaces,
then the above definition coincides with the usual definition of
computational model in which $\phi$ defines a homeomorphism from
$(X,\tau)$ onto the space of maximal elements of $(P,\sqsubseteq)$
\cite{kop-kunz,Mart-Mis-Reed}.

Next, we define the notion of a quasi-metric space. For more details
the reader may consult the references
\cite{fletcher,kelly,nonsymmetric,leftkunzi,kunzi-schell}. A {\em
quasi-metric} $d$ on a set $X$ is a function $d:X\times X\rightarrow
[0,\infty)$ such that for any $x,y,z\in X$:
\begin{enumerate}[(1)]
\item $x=y$ iff $d(x,y)=d(y,x)=0$,
\item $d(x,z)\leq d(x,y)+d(y,z)$.
\end{enumerate}
If we drop the if part of condition (1), $d$ is called a {\em
quasi-pseudometric}. The pair $(X,d)$ is called
quasi-(pseudo)metric space. Each quasi-metric $d$ on the set $X$
induces a $T_0$ topology on $X$, denoted by $\tau_d$, whose base
is the set of all balls of the form ${\mathcal N}_\epsilon(x)=\{y\in
X:\ d(x,y)<\epsilon\}$, for any $x\in X$ and $\epsilon>0$. The
topology $\tau_d$ is $T_1$ if and only if the condition (1) can
be replaced by: $x=y\ \Leftrightarrow\ d(x,y)=0$. The
quasi-metric $d$ generates another quasi-metric $d^{-1}$ on the
set $X$, called the {\em conjugate} of $d$, defined by
$d^{-1}(x,y)=d(y,x)$. Also, the function $d^*$ can be defined on
$X\times X$ by $d^*(x,y)=max\{d(x,y),d^{-1}(x,y)\}$ which is a
metric on $X$. The quasi-metric space $(X,d)$ is {\em point
symmetric} if $\tau_d\subseteq \tau_{d^{-1}}$. For example, any
compact $T_1$ quasi-metric space $(X,d)$ is point symmetric
(\cite{weston}, Lemma 2). A sequence $(x_n)_{n>0}$ is called {\em
Cauchy} ({\em biCauchy}) sequence if for every $\epsilon>0$ there
is $N>0$ such that $d(x_n,x_m)<\epsilon$ whenever $m\geq n\geq N$
($m,n\geq N$). An element $x\in X$ is called a {\em Yoneda limit}
of the sequence $(x_{n})_{n>0}$, if for any $y\in X$,
\[
d(x,y)=\inf_{n}\sup_{m\geq n} d(x_{m},y).
\]
The quasi-metric space $(X,d)$ is {\em sequentially Yoneda-complete}
if every Cauchy sequence has a Yoneda limit. It is easy to see that
the Yoneda limit is unique if it exists. A point $e\in X$ is called
{\em finite } if for any Cauchy sequence $(x_{n})_{n>0}$
in $X$ with the Yoneda limit $x$,
\[
d(e,x)=\sup_{n}\inf_{m\geq n} d(e,x_{m}).
\]
The quasi-metric space $(X,d)$ is called {\em algebraic} if each
element of $X$ is the Yoneda limit of a Cauchy sequence of finite
elements. The quasi-metric space $(X,d)$ is {\em Smyth-complete}
if any Cauchy sequence $(x_{n})_{n>0}$ converges strongly in $X$,
i.e.\ there is a point $x\in X$ such that $(x_{n})_{n>0}$
converges to $x$ in the topology of the metric $d^*$.

Finally, we review some basic definitions from the hyperspace
theory \cite{cao2,rodri-rom}. Let $(X,d)$ be a bounded
quasi-metric space and $\mathcal{K}_0(X)$ denote the set of all
nonempty compact subsets of $X$. The upper Hausdorff
quasi-pseudometric $H^+_d$ and the lower Hausdorff
quasi-pseudometric $H^-_d$ on $\mathcal{K}_0(X)$ are defined as
follows:
\[
H^+_d(A,B)= \sup_{b\in B}d(A,b)\ \ \ \ \ , \ \ \ \ \
H^-_d(A,B)=\sup_{a\in A}d(a,B)
\]
for all $A,B\in\mathcal{K}_0(X) $, where $d(A,x)=\inf_{a\in
A}d(a,x)$ and $d(x,A)=\inf_{a\in A}d(x,a)$. The Hausdorff
quasi-pseudometric $H_d$ is defined as $H^+_d \vee H^-_d$ or
equivalently
\[
H_d(A,B)= \max\{\sup_{b\in B}d(A,b),\sup_{a\in A} d(a,B)\}
\]
for all $A,B\in\mathcal{K}_0(X)$. It is known that $H^+_d$, $H^-_d$
and $H_d$ are quasi-pseudometrics on $\mathcal{K}_0(X)$. For a $T_1$
quasi-metric space $(X,d)$, $H_d$ is a quasi-metric. Furthermore,
for any $A,B\in \mathcal{K}_0(X)$,
\[
\hspace{3cm}H_d(A,B)=0\;\;\;\;\;\; \mbox{if and only if}\ \;\;\;\;\;\;\
B\subseteq A\subseteq cl_{\tau_d}{B}.\hspace{3.3cm} (*)
\]
In \cite{3}, the authors present an example which shows that
$(\mathcal{K}_0(X),H_d)$ may not be a $T_1$ space, even though
$(X,d)$ is a $T_1$ quasi-metric space. However, one can infer from
$(*)$ that $(\mathcal{K}_0(X),H_d)$ is $T_1$ if $(X,d)$ is
Hausdorff (more generally KC-space in which all compact subsets are
closed).

Recall that a subset $K$ of a quasi-metric space $(X,d)$ is
$d$-precompact if for any $\epsilon>0$, there is a finite subset
$F$ of $K$ such that for any $k\in K$, $d(x,k)<\epsilon$, for some
$x\in F$. Unlike the metric spaces, a compact subset of a
quasi-metric space $(X,d)$ is not necessarily
$d^{-1}$-precompact. The following theorem shows that if we
impose this extra condition to $(X,d)$, then the
Smyth-completeness of $(X,d)$ can be lifted up to
$(\mathcal{K}_0(X),H_d)$. This  theorem  is used in section 5,
Lemma \ref{biCauchy}.

\begin{thm}\label{smythcomplete}
Let $(X,d)$ be a Smyth-complete quasi-metric space such that any compact
subset of $X$ is $d^{-1}$-precompact. Then $(\mathcal{K}_0(X),H_d)$
is Smyth-complete.
\end{thm}
\proof See  \cite{3}, Theorem 3.7.\qed

There are other topologies on the hyperspace $\mathcal{K}_0(X)$.
The most famous of these topologies is the Vietoris topology
$\tau_V$ which is the supremum of lower Vietoris topology and
upper Vietoris topology. The lower Vietoris topology
${\mathcal\tau}^-_V$ is generated by all sets of the form
$\diamondsuit V=\{K\in\mathcal{K}_0(X):\ K\cap V\neq \emptyset\}$
whereas the upper Vietoris topology ${\mathcal\tau}^+_V$ is generated
by all sets of the form $\Box V=\{K\in\mathcal{K}_0(X):\
K\subseteq V\}$ for open $V$. In general, this topology is coarser than the
topology of Hausdorff quasi-pseudometric $H_d$ on the hyperspace
$\mathcal{K}_0(X)$. However, whenever any compact subset of $X$ is
$d^{-1}$-precompact, these topologies coincide \cite{rodri-rom}.

\section{The space of formal balls and its $\omega$-Plotkin domain}

The space of formal balls of a metric space $(X,d)$, denoted by
$\mathbf{B}X$, was defined by Edalat and Heckmann in \cite{edalat}.
This construction gives a concrete computational model for metric
spaces in which the order-theoretic properties of $\mathbf{B}X$ are
closely connected with the metric properties of $(X,d)$.

In \cite{Rutten}, Rutten was probably the first who studied the
space of formal balls for quasi-metric spaces via co-Yoneda
embedding. More recently, Ali-Akbari et al. in \cite{1} and
Romaguera and Valero in \cite{Rom-Val,Rom-Val1} studied the set of
formal balls for quasi-metric spaces in the spirit of Edalat and
Heckmann's work.

\begin{defi}\label{dfnfor}
For a quasi-metric space $(X,d)$, the space of {\em formal balls}
is the pair $(\mathbf{B}X,\sqsubseteq)$ where
\[
\mathbf{B}X=\{(x,r):\ x\in X\ \mbox{and}\ r\geq 0\},
\]
and
\[
(x,r)\sqsubseteq (y,s)\;\;\;\;\;\  \mbox{if and only if}\;\; \;\;\
d(x,y)\leq r-s,
\]
for any $(x,r),(y,s)\in \mathbf{B}X$. It is easy
to see that $(\mathbf{B}X,\sqsubseteq)$ is a poset. An element
$(x,r)$ of $\mathbf{B}X$ is called a formal ball. One can define
an auxiliary relation $\prec$ on $\mathbf{B}X$ as follows:
\[
(x,r)\prec(y,s)\;\;\;\;\;\  \mbox{if and only if}\;\;\;\;\ d(x,y)< r-s.
\]
It can be shown that the relation $\prec$ satisfies the interpolation
property and therefore $(\mathbf{B}X,{\prec})$ forms an abstract basis.
\end{defi}

The following theorem shows some interesting properties of the poset of
formal balls.

\begin{thm}\label{2}
Let $(X,d)$ be a quasi-metric space.
\begin{enumerate}[\em(1)]
\item The function $\iota:(X,\tau_d)\rightarrow (\mathbf{B}X,\mathbf{P}\sigma)$ defined by $\iota(x)=(x,0)$
is an embedding.
\item If $(X,d)$ is $T_1$ and sequentially Yoneda-complete, then $(\mathbf{B}X,\sqsubseteq)$ is a
dcpo. In addition, if $(X,d)$ is also algebraic,
$(\mathbf{B}X,\sqsubseteq)$ is a domain (continuous dcpo).
\item If $(X,d)$ is Smyth-complete, then $(\mathbf{B}X,\sqsubseteq)$ is a
domain. Moreover, the auxiliary relation $\prec$ coincides with the
way-below relation of $\mathbf{B}X$.
\end{enumerate}
\end{thm}
\proof See Theorem 3.12, Corollary 3.13 and Theorem 3.17 in \cite{1}. \qed

In the light of the first part of the above Theorem, to ease our
notations, we identify the set $X$ with the set
$\iota(X)=\{(x,0):\ x\in X\}$. In particular, any compact subset
of $X$ is identified with a compact subset of the set $\iota(X)$.

Now, we review the definition of chain completion of an abstract
basis $(P,\prec)$. The set of all $\omega$-chains, i.e.
$\prec$-ascending sequences, of $P$ is denoted by $CP$.

\begin{defi}
For two $\omega$-chains $(x_n)_{n>0}$ and $(y_m)_{m>0}$ in $CP$ define
\[
(x_n)_{n>0}\sqsubseteq(y_m)_{m>0}\
\Leftrightarrow\ \forall n \exists m \ \  x_n \prec y_m,
\]
\[
(x_n)_{n>0}\sim(y_m)_{m>0}\ \Leftrightarrow\
(x_n)_{n>0}\sqsubseteq(y_m)_{m>0}\ \& \
(y_m)_{m>0}\sqsubseteq(x_n)_{n>0}.
\]
The {\em chain completion} of the abstract basis $(P,\prec)$ is
defined to be the partially ordered set $(\mathcal{C}P,\sqsubseteq)$
where $\mathcal{C}P=CP/{\sim}$ and
\[
[(x_n)_{n>0}]\sqsubseteq[(y_m)_{m>0}]\
\Leftrightarrow\ (x_n)_{n>0}\sqsubseteq(y_m)_{m>0},
\]
for any $[(x_n)_{n>0}]$ and $[(y_m)_{m>0}]$ in $\mathcal{C}P$. It
is a well-known fact that $(\mathcal{C}P,\sqsubseteq)$ is a
continuous $\omega$-dcpo \cite{knijnenburg}. The way-below
relation is given by:
\[
[(x_n)_{n>0}]\ll[(y_m)_{m>0}]\
\Leftrightarrow\ \exists m \forall n\ \ x_n\prec y_m.
\]
$(x_n)_{n>0}$ is called a {\em representation} of
$[(x_n)_{n>0}]\in{\mathcal{C}P}$.  By abuse of notation,
for any equivalence class $\mathcal I$ of ${\mathcal{C}P}$, we write
$x\in {\mathcal I}$ if $x$ is an element of one of the sequences
representing $\mathcal I$.
\end{defi}

\begin{defi}\label{Egli-Milner}
For subsets $A$ and $B$ of the abstract basis $(P,\prec)$, define
\begin{enumerate}[(1)]
\item\hspace{1.5cm} $A\prec_U B \ \Leftrightarrow \ \forall \ b\in B \ \exists \ a\in A \ \ a\prec b$,
\item\hspace{1.5cm} $A\prec_L B \ \Leftrightarrow \ \forall \ a\in A \ \exists \ b\in B \  \ a\prec b$,
\item\hspace{1.5cm} $A\prec_{EM}B\ \ \mbox{if and only if}\ \ A\prec_U B\ \mbox{and}\ A\prec_LB$.
\end{enumerate}
The relation $\prec_{EM}$ stands for Egli-Milner relation. Let
$\mathcal{P}_{fin}P$ be the set of all non-empty finite subsets of
$P$. It is easy to see that $(\mathcal{P}_{fin}P,\prec_{EM})$ is
an abstract basis.  Since there is no danger of confusion, for
brevity, we drop the subscript $EM$. The chain completion of
$(\mathcal{P}_{fin}P,\prec)$ is called {\em $\omega$-Plotkin
domain} of $P$ which is denoted by ${\bf C}P$.
\end{defi}

In particular, for a quasi-metric space $(X,d)$, we consider the
$\omega$-Plotkin domain ${\bf C}{\bf B}X$ of the abstract basis $({\bf
B}X,\prec)$.  For
$F\in \mathcal{P}_{fin}{\bf B}X$ and ${\mathcal I}\in {\bf C}{\bf B}X$,
define
\[
rF=\max\{r:\ (x,r)\in F\}\ \ \mbox{and}\ \ \overline{r}{\mathcal
I}=\inf\{rF:\ F\ \mbox{belongs to a representation of}\ {\mathcal
I}\}.
\]
Note that ${\mathcal I}\sqsubseteq {\mathcal J}$ implies that
$\overline{r}{\mathcal I}\geq \overline{r}{\mathcal J}$. Also from
$F\prec G$,  $rF>rG$ follows. Therefore, if ${\mathcal
I}\ll{\mathcal J}$ then $\overline{r}{\mathcal I}> \overline{r}{\mathcal J}$.

Below,  an important property of this structure is highlighted.
\begin{thm}\label{dcpo}
Let $(X,d)$ be a quasi-metric space. Then $\omega$-Plotkin domain
$({\bf C}{\bf B}X,\sqsubseteq)$ is a continuous $\omega$-dcpo and moreover,
any $\ll$-directed subset of it has a least upper bound.
\end{thm}
\proof As we mentioned earlier, the first part is known.
For the second part, let $\preceq$ be the partial order relation
generated   by $\ll$, i.e.
\[
\mathcal{I}\preceq \mathcal{J}\;\;\;\;\  \mbox{ if and only if}\;\;\;\;\;
\mathcal{I}=\mathcal{J}\ \mbox{or}\ \mathcal{I}\ll \mathcal{J},
\]
for every $\mathcal{I},\mathcal{J}\in {\bf C}{\bf B}X$.

We first  verify that $({\bf C}{\bf B}X, \preceq)$ is a dcpo. By
a well-known fact from \cite{Mar} it suffices to examine  that
every $\preceq$-chain of ${\bf C}{\bf B}X$ has a least upper
bound. Let ${\mathcal A}=({\mathcal I}_\alpha)_{\alpha\in I}$ be a
$\preceq$-chain in ${\bf C}{\bf B}X$. Without loss of generality,
we may assume that ${\mathcal A}$ has no maximum element. Then $(
\overline{r}{\mathcal I}_\alpha)_{\alpha\in I}$ is a strictly
decreasing chain in the set  of nonnegative real numbers and
therefore has an infimum, say $r$. Fix ${\mathcal I}_1$ and
inductively for any $n\geq 2$, choose ${\mathcal I}_{n-1}\ll{\mathcal
I}_n$ such that $r<\overline{r}{\mathcal I}_n<r+\frac{1}{n}$. We
claim that  $({\mathcal I}_n)_{n>0}$ is a cofinal subsequence of
$\mathcal A$ in ${\bf C}{\bf B}X$. Let ${\mathcal I}_{\alpha}$ be an
arbitrary element of $\mathcal A$. Then choose ${\mathcal I}_n$ in such a
way that $\overline{r}{\mathcal I}_n < \overline{r}{\mathcal I}_\alpha$.
Now since any two elements of $\mathcal A$ are comparable and
$\overline{r}{\mathcal I}_n < \overline{r}{\mathcal I}_\alpha$, it
follows that ${\mathcal I}_\alpha \ll {\mathcal I}_n$. But $({\bf C}{\bf
B}X,\sqsubseteq)$ is an $\omega$-dcpo. Therefore, $({\mathcal
I}_n)_{n>0}$ has the least upper bound $\mathcal I$ in $({\bf C}{\bf
B}X,\sqsubseteq)$, which is also the least upper bound  for $\mathcal
A$.

Now, let $\mathcal D$ be a $\ll$-directed subset  of ${\bf C}{\bf
B}X$. Then it is easy to see that $\mathcal D$ is also
$\preceq$-directed. Therefore, it has the $\preceq$-least upper
bound $\mathcal I$. It is, then,  straightforward to show that $\mathcal
I$ is also the $\ll$-least upper bound of $\mathcal D$. \qed

Although we are not able to show that ${\bf C}{\bf B}X$ is a
dcpo, the above Theorem gives a crucial feature of ${\bf C}{\bf
B}X$ which will help us in obtaining Theorem \ref{domrep}.

We introduce the following abbreviations which will be useful
for a number of later proofs.

\begin{note}\label{note}
Let $F,G\in \mathcal{P}_{fin}{\bf B}X$ and $\epsilon>0$ be given.

\begin{enumerate}[(1)]
\item  $F+\epsilon =\{(x,r+\epsilon):\ (x,r)\in F\}$.
\item  If  $F\prec G$, then put $\delta(F,G)=\min\{(r-s)-d(x,y)\ :\ (x,r)\in F \ , \ (y,s)\in G\ ,
\ (x,r)\prec (y,s)\}$.
\end{enumerate}
\end{note}

\begin{rem}\label{rem1}
The following properties of the above notations are  straightforward.

\begin{enumerate}[(1)]

\item For $0<\epsilon'<\epsilon$, $F+\epsilon\prec F+\epsilon'\prec F$.

\item For any $\epsilon<\delta(F,G)$, $(x,r)\in F$ and $(y,s)\in G$, with
$(x,r)\prec (y,s)$, we have $(x,r)\prec (y,s+\epsilon)$. Hence $F\prec G+\epsilon$.

\end{enumerate}
\end{rem}

\section{Embedding of ${\mathcal K}_0(X)$ into ${\bf C}{\bf B}X$}

In this section, we apply the techniques used by Edalat and
Heckmann in \cite{edalat} to the $\omega$-Plotkin domain of the
space of formal balls, leading us to find a computational model
for the space ${\mathcal K}_0(X)$ of the nonempty compact subsets of a
quasi-metric space $(X,d)$. More precisely, for a  sequentially
Yoneda-complete $T_1$ quasi-metric space $(X,d)$, we construct
the $\omega$-Plotkin domain of the abstract basis $({\bf
B}X,\prec)$, as introduced in Definition \ref{dfnfor}, and show
that the hyperspace ${\mathcal K}_0(X)$ equipped with the Vietoris
topology $\tau_V$ can be embedded in ${\bf C}{\bf B}X$ equipped
with the Scott topology. Moreover, this embedding serves as an
$\omega$-computational model of  $({\mathcal K}_0(X),\tau_V)$.

From now on, we assume that $(X,d)$ is a sequentially Yoneda-complete $T_1$
bounded quasi-metric space.

\begin{defi}
Let $K$ be a nonempty compact subset of $X$. Since $K$ is compact,
it is $d$-precompact. So, for any $n>0$, one can choose
$x_1^n,\ldots ,x_{m_n}^n$ of $K$ such that for any $x\in K$,
$d(x^n_i,x)<\frac{1}{2n^2}$ for some $x_i^n$. Put
\[
F_n=\{(x_1^1,\frac{1}{n}),\ldots,(x_{m_1}^1,\frac{1}{n}),\ldots,(x_1^n,\frac{1}{n}),\ldots,(x_{m_n}^n,\frac{1}{n})\}.
\]
It can be easily checked that $( F_n)_{n>0}$ is an
$\omega$-chain in $({\mathcal P}_{fin}({\bf B}X),\prec)$. Call $(
F_n)_{n>0}$  a {\em standard } representation of $K$.
\end{defi}

Although a compact subset $K$ of $X$ might have several standard
representations,  we show that all standard
representations of $K$ are $\sim$-equivalent. The following
auxiliary lemmas will be useful in several proofs.

\begin{lem}\label{way}
For any $x\in K$ and any standard representation $(
F_n)_{n>0}$ of $K$, there is a sequence $(
(x_n,r_n))_{n>0}$ such that $(x_n,r_n)\in F_n$ and
$d^*(x_n,x)\rightarrow 0$.
\end{lem}
\proof First note that for any $n>0$, $F_n \prec K$. Hence
for any $n>0$, there is $(x_n,r_n)\in F_n$ such that
$(x_n,r_n)\prec(x,0)$. As $rF_n\rightarrow 0$, it implies
$d(x_n,x)\rightarrow 0$. Now, since $K$ is compact and therefore
$d$ is point symmetric on $K$, it follows $d(x,x_n)\rightarrow
0$. Therefore $d^*(x_n,x)\rightarrow 0$.\qed

Let $\mathcal I$ be an element of ${\bf C}{\bf B}X$. We can
obtain a nonempty compact saturated subset of ${\bf B}X$, denoted by
${\mathcal I}^+$, as \[{\mathcal I}^+=\bigcap_{n>0}\uparrow
F_n,\] where $( F_n)_{n>0}$ is a representation of $\mathcal
I$. In fact, since ${\mathcal I}^+$ is a filtered intersection of
nonempty compact saturated subsets of $({\bf B}X,\mathbf{P}\sigma)$
and ${\bf B}X$ is a dcpo, similar to Theorem 7.2.27 in
\cite{abramsky}, it can be proved that the pseudoScott topology
$\mathbf{P}\sigma$ is sober. So ${\mathcal I}^+$ is a nonempty compact
saturated subset of ${\bf B}X$. Also, ${\mathcal I}^+$ is independent of
the choice of its representations. One can easily show that for any
representations $( F_n)_{n>0}$ and $(
E_m)_{m>0}$ of $\mathcal I$,
\[
\bigcap_n\uparrow F_n=\bigcap_m\uparrow E_m.
\]

\begin{lem}\label{1}
Let $\mathcal{I}$ be in ${\bf C}{\bf B}X$ with $\overline{r}{\mathcal
I}=0$. Then for any standard representation $( F_n)_{n>0}$ of
$\iota^{-1}({\mathcal I}^+)$ and any $G,H\in\mathcal{P}_{fin}{\bf B}X$ with
$G\prec H\prec {\mathcal I}^+$, $G\sqsubseteq F_n$ for some $n>0$.
\end{lem}
\proof Let $(y,s)\in G$. Since $G\prec{\mathcal I}^+$, there is
$(x,0)\in{\mathcal I}^+$ such that $(y,s)\prec (x,0)$. Now as
$(x,0)\in \bigcap_{n>0}\uparrow F_n$, by Lemma \ref{way}, there is
a sequence $( (x_n,r_n))_{n>0}$ such that $(x_n,r_n)\in F_n$ and
$d^*(x_n,x)\rightarrow 0$. For $\epsilon=s-d(y,x)$, take $n>0$
such that $d(x,x_n)<\epsilon/2$ and $r_n<\epsilon/2$. One can
readily see that $(y,s)\prec(x_n,r_n)$. So, for any $(y,s)\in G$,
there is $F_n$ and $(x_n,r_n)\in F_n$ such that
$(y,s)\prec(x_n,r_n)$. Assume that $F$ in $( F_n)_{n>0}$ is an
upper bound for all $F_n$, arisen in this way. Now,
$G\sqsubseteq_L F$ is straightforward.

Next, put $\delta=\delta(G,H)$, as defined in Notation
\ref{note}. Choose $F_n$ such that $rF_n<\delta$ and  $F \prec
F_n$. Take $(x,r)\in F_n$.  Then $(x,0)\in {\mathcal I}^+$.
Therefore, there exists $(z,t)\in H$ such that $(z,t)\prec(x,0)$.
Also, $G\prec_U H$ implies that there exists $(y,s)\in G$ such
that $(y,s)\prec(z,t)$. It follows that
\[
\begin{array}{ll}
d(y,x)&\leq d(y,z)+d(z,x)\\
&\leq (s-t)-\delta+t\\
&<s-r.
\end{array}
\]
Hence $G\prec_U F_n$. Now from $G\sqsubseteq_L F$ and $F\prec
F_n$, it follows that $G\sqsubseteq_L F_n$ and consequently  $G
\sqsubseteq F_n$.\qed

\begin{prop}\label{well-defined}
For any $K\in{\mathcal K}_0(X)$, all standard representations of $K$
are $\sim$-equivalent.
\end{prop}
\proof Let $( F_n)_{n>0}$ and $( E_m)_{m>0}$ be two
standard representations of $K$. For any $m>0$, since $E_m\prec
E_{m+1}\prec K$, it follows from Lemma \ref{1} that there is
$n>0$ such that  $E_m \prec F_n$. Therefore $( E_m)_{m>0}
\sqsubseteq ( F_n)_{n>0}$. Similarly, it can be proved that $(
F_n)_{n>0}\sqsubseteq( E_m)_{m>0}$ and therefore $( E_m)_{m>0}$
and $( F_n)_{n>0}$
are in an equivalence class. \qed

In the light of the above proposition, the following definition
is established.

\begin{defi}
Denote the equivalence class of a standard representation $(
F_n)_{n>0}$ of a nonempty compact subset $K$ by $K^*$ and
let $\phi:{\mathcal K}_0(X)\rightarrow{\bf C}{\bf
B}X$ be  $\phi(K)=K^*$.
\end{defi}

We prove some properties of this map.

\begin{prop}\label{max}
Let $K,L\in {\mathcal K}_0(X)$ and ${\mathcal I}\in {\bf C}{\bf B}X$ with
$\overline{r}{\mathcal I}=0$. Then
\begin{enumerate}[\em(1)]
\item ${K}=(K^*)^+$.

\item For any representation $( F_n)_{n>0}$ of $\mathcal I$;
\[
{\mathcal I}^+=\{\bigsqcup_n a_n\ :\ a_n\in F_n\ ,\ (a_n)_{n>0}\ \mbox{is an ascending sequence}\}.
\]

\item $K^*\sqsubseteq L^*$ implies $L\subseteq K\subseteq cl_{\tau_d}L$,
where $cl_{\tau_d}L$ is the closure of $L$ in $\tau_d$.
\end{enumerate}
\end{prop}
\proof (1) It is routine to check that ${K}\subseteq
(K^*)^+$. For the opposite inclusion, suppose that there is
$(x,0)\in (K^*)^+\setminus {K}$. For any $(y,0)\in {K}$, put
$s_y=\frac{1}{2}d(y,x)$. Using compactness of $K$, choose a finite
subset $G_0$ of $\{(y,\frac{1}{2}s_y):\ (y,0)\in {K}\}$ such that
$G_0\prec {K}$. Select $F\in K^*$ with $rF<\min \{\frac{1}{2}s_y\
:\ (y,\frac{1}{2}s_y)\in G_0\}$. We claim that $G\prec_U F$,
where $G=\{(y,s_y):(y,\frac{1}{2}s_y)\in G_0\}$. To prove the
claim, let $(z,t)\in F$. So $(z,0)\in {K}$ and therefore there is
$(y,\frac{1}{2}s_y)\in G_0$ such that $(y,\frac{1}{2}s_y)\prec
(z,0)$. Since $t<\frac{1}{2}s_y$, it follows $(y,s_y)\prec(z,t)$
and consequently $G\prec_U F$.

Now, by definition of $(K^*)^+$, it is clear that $F\prec_U
(K^*)^+$. So there is $(a,u)\in F$ with $(a,u)\prec (x,0)$. Since
$G\prec_U F$, there is $(y,s_y)$ in $G$ such that $(y,s_y)\prec
(a,u)$. Hence $(y,s_y)\prec (x,0)$ or equivalently $d(y,x)<s_y$,
which is a
contradiction.\\

(2) Clearly the supremum of any ascending sequence $(a_n)_{n>0}$,
where $a_n\in F_n$, belongs to ${\mathcal I}^+$. Let
$(a,0)\in {\mathcal I}^+$. Put $G_n=\{(x,r)\in F_n:\ (x,r)\prec
(a,0)\}$. For any $(x,r)$ and $(y,s)$ in $\bigcup_nG_n$, define
$(x,r)R(y,s)$ if and only if for some $n>0$, $(x,r)\in
G_n$, $(y,s)\in G_{n+1}$ and $(x,r)\prec(y,s)$. The binary relation $R$ defines
a locally finite directed graph on the infinite set $G=\bigcup_nG_n$ with at most
$|F_1|$-connected components. So  $G$ has an infinite connected component and therefore,
in the light of K\"onig's Lemma, there is an ascending sequence
$((x_n,r_n))_{n>0}$ such that $\bigsqcup_n
(x_n,r_n)=(b,0)\sqsubseteq (a,0)$. Since $d$ is $T_1$,
$b=a$ follows and the proof is complete.\\

(3) The assumption implies $(L^*)^+\subseteq (K^*)^+$. So in the
light of the first part, $L\subseteq K$. For $K \subseteq
cl_{\tau_d}L$, let $x\in K$. Suppose that $( F_n)_{n>0}$ and $(
G_n)_{n>0}$ are standard representations of $K$ and $L$,
respectively. By the second part, there is an ascending sequence
$( (x_n,r_n))_{n>0}$, $(x_n,r_n)\in F_n$, such that $\bigsqcup_n
(x_n,r_n)=(x,0)$. So $d(x_n,x)\rightarrow 0$. Since $K$ is
compact and therefore $d$ is point symmetric on $K$,
$d(x,x_n)\rightarrow 0$. From $K^*\sqsubseteq L^*$, it follows
that for any $F_n$, there is $G_{m_n}$  and $(y_{m_n},s_{m_n})\in
G_{m_n}$ such that $F_n\sqsubseteq G_{m_n}$ and $(x_n,r_n)\prec
(y_{m_n},s_{m_n})$. Thus $d(x,y_{m_n})\rightarrow 0$, which means
$x\in cl_{\tau_d}L$.\qed

From the first part of the above proposition, it follows that the
map $\phi:{\mathcal K}_0(X)\rightarrow {\bf C}{\bf B}X$ is one-to-one.
Moreover, in the following we prove that the map $\phi$ gives a
one-to-one correspondence between the maximal element of ${\mathcal
K}_0(X)$ with respect to the specialization order $\sqsubseteq_{H_d}$
and the maximal element of the partially ordered set ${\bf C}{\bf B}X$.
Before proving this, we need the following lemma.

\begin{lem}\label{max1}
For any maximal element $\mathcal{I}$ of ${\bf C}{\bf B}X$,
$\mathcal{I}=I^*$ where $I=\iota^{-1}({\mathcal I}^+)$.
\end{lem}
\proof Note that ${\mathcal I}^+$ is a nonempty compact subset of ${\bf
B}X$. Since $\iota: (X,\tau_d)\rightarrow ({\bf
B}X,\mathbf{P}\sigma)$ is an embedding (Theorem \ref{2}), $I=\iota^{-1}({\mathcal
I}^+)$ is a compact set in $(X,\tau_d)$. Thus $I^*$ is well-defined.

Now, by maximality of $\mathcal I$, it suffices to prove that
$\mathcal{I}\sqsubseteq I^*$. Let $( F_n)_{n>0}$ be a
representations of $\mathcal I$ and $( G_m)_{m>0}$ be a
standard representation of $I$. Take $F_n$ in $( F_n)_{n>0}$.
Since $F_n\prec F_{n+1}\prec{\mathcal I}^+$, according to
Lemma \ref{1}, $F_n\sqsubseteq G_m$ for some $m>0$. This
shows $\mathcal{I}\sqsubseteq I^*$.\qed

\begin{prop}\label{maximal}
For any maximal element $K$ of ${\mathcal K}_0(X)$ with respect to the
specialization order $\sqsubseteq_{H_d}$, $K^*$ is maximal in $({\bf
C}{\bf B}X,\sqsubseteq)$. Conversely, any maximal element $\mathcal I$
of ${\bf C}{\bf B}X$ is of the form $K^*$ for some maximal element
$K$ of $({\mathcal K}_0(X),\sqsubseteq_{H_d})$.
\end{prop}
\proof Let $K$ be a maximal element of $({\mathcal
K}_0(X),\sqsubseteq_{H_d})$ and $K^*\sqsubseteq {\mathcal I}$. Without
loss of generality, we assume that $\mathcal I$ is maximal. So ${\mathcal
I}=I^*$, where $I=\iota^{-1}({\mathcal I}^+)$. We have $K^*
\sqsubseteq I^*$ and therefore by the third part of Proposition
\ref{max}, $I\subseteq K\subseteq cl_{\tau_d}I$. Hence
$H_d(K,I)=0$. By maximality of $K$, we conclude that $K=I$ and
$K^*={\mathcal I}$.

Now, let $\mathcal I$ be a maximal element of ${\bf C}{\bf B}X$. Lemma
\ref{max1} implies that for any maximal element $\mathcal I$ of ${\bf
C}{\bf B}X$, $\mathcal{I}=I^*$, where $I=\iota^{-1}({\mathcal I}^+)$
is a nonempty compact subset of $(X,d)$. To complete the proof, we
have to show that $I$ is maximal in $({\mathcal
K}_0(X),\sqsubseteq_{H_d})$. Let $I\sqsubseteq_{H_d}J$ for some
$J\in {\mathcal K}_0(X)$. Thus $H_d(I,J)=0$ and consequently
$J\subseteq I\subseteq cl_{\tau_d}J$. We show that
$I^*\sqsubseteq J^*$. Suppose $( F_n)_{n>0}$ and $( G_m)_{m>0}$
are standard representations of $I$ and $J$, respectively. Since
$J\subseteq I\subseteq cl_{\tau_d}J$, it follows that $F_n\prec
F_{n+1}\prec J$, for any $n>0$. By Lemma \ref{1}, $F_n\sqsubseteq
G_m$ for some $m>0$. So $I^*\sqsubseteq J^*$ and by maximality of
$I^*$ it follows that $I=J$. Therefore $I$ is maximal in $({\mathcal
K}_0(X),\sqsubseteq_{H_d})$.\qed

Below we examine different topologies on ${\mathcal K}_0(X)$ for which
the map $\phi$ becomes an embedding.

\begin{thm}\label{vietoris}
The map $\phi$ is an embedding from the hyperspace ${\mathcal K}_0(X)$
equipped with the Vietoris topology into ${\bf C}{\bf B}X$ with
the Scott topology.
\end{thm}
\proof Let $\Uun {\mathcal I}=\{{\mathcal J}\in {\bf C}{\bf B}X\ :\
{\mathcal I}\ll{\mathcal J} \}$ be a basic open set of ${\bf C}{\bf B}X$
in the Scott topology and $K\in \phi^{-1}(\Uun {\mathcal I})$.
So $K^*\in \Uun {\mathcal I}$ or equivalently ${\mathcal I}\ll K^*$. Let $(
F_n)_{n>0}$ be a standard representation of $K^*$. By definition
of the way-below relation, there is $N>0$ such that for any element
$G\in {\mathcal I}$, $G \prec F_{N-1}$. Define an open subset $\mathcal V$
of $({\mathcal K}_0(X),{\mathcal \tau}_{V})$ as ${\mathcal
V}=(\bigcap_{(x,r)\in F_N}\diamondsuit V_x)\cap \Box V$, where
$V_x={\mathcal{N}}_{\frac{1}{N}}(x)$ and $V=\bigcup_{(x,r)\in
F_N}{\mathcal{N}}_{\frac{1}{N}}(x)$. Clearly $K\in {\mathcal V}$. We show
that ${\mathcal V}\subseteq \phi^{-1}(\Uun {\mathcal I})$. Let $B\in {\mathcal
V}$. First, we prove that $F_N\prec {B}$. Take $(x,\frac{1}{N})\in
F_N$. From $B\in \diamondsuit V_x$, it follows that there is
$b\in B\cap V_x$. Thus $d(x,b)<\frac{1}{N}$ or equivalently
$(x,\frac{1}{N})\prec (b,0)$ and therefore $F_N\prec_L{B}$.
Finally $F_N\prec_U{B}$ follows from $B\in \Box V$. Now, by Lemma
\ref{1}, for any standard representation of $B$, there is an
element $H$ in this representation such that $F_{N-1}\prec H$. So
for any element $G\in {\mathcal I}$, $G\prec H$. That means ${\mathcal
I}\ll B^*$.

On the other hand, we prove that the image of any upper (resp.
lower) Vietoris open subset under the map $\phi$ is open in the
relative Scott topology on $\phi({\mathcal K}_0(X))$. Let $\Box V$ be
an upper Vietoris open subset of $\mathcal{K}_0(X)$ and $K^*\in
\phi(\Box V)$. Take $\epsilon>0$ such that $\bigcup_{x\in K}
\mathcal{N}_\epsilon(x)\subseteq V$. Assume that $( F_n)_{n>0}$ is
a standard representation of $K$. Choose $N>0$ such that
$rF_N<\epsilon$ and put ${\mathcal I}_K= [(E_n)_{n>0}]$, where
$E_n=F_N+\frac{1}{n}$ (See Notation \ref{note}.). Remark
\ref{rem1}(1), indicates that the sequence  $(E_n)_{n>0}$ is
$\prec$-ascending and therefore  $(E_n)_{n>0}$ is an
$\omega$-chain. Clearly ${\mathcal I}_K\ll K^*$. We prove
\[
\hspace{5.5cm}\Uun {\mathcal I}_K\cap \phi({\mathcal K}_0(X))\subseteq
\phi(\Box V).\hspace{5cm}(*)
\]
Let $B^*\in \Uun{\mathcal I}_K$. We show that $B^*\in \phi(\Box V)$ or
equivalently $B\subseteq V$. Let $b\in B$. Because of ${\mathcal I}_K\ll
B^*$, there is an element $H\in B^*$ such that $E_n\prec H$, for any
$n>0$. So there are $(z,t)\in H$ and $(x_n,r_n+1/n)\in
E_n$, for any $n>0$, with
$(x_n,r_n+1/n)\prec(z,t)\prec(b,0)$. Since the set $\{(x_n,r_n)\ :\
(x_n,r_n+1/n)\prec(b,0)\}\subseteq F_N$ is finite, it follows that
there is $(x,r)\in F_N$ such that $d(x,b)\leq r<\epsilon$. Thus from
$x\in K$,  $b\in V$ follows and consequently $B^*\in \phi(\Box V)$.

Now, let $\diamondsuit V$ be a lower Vietoris open subset of
$\mathcal{K}_0(X)$. Suppose $K^*\in \phi(\diamondsuit V)$ and $(
F_n)_{n>0}$ is a standard representation of $K^*$.
Select $x\in K\cap V$ and put $F'_n=F_n\cup\{(x,1/n)\}$. One
can readily see that $( F'_n)_{n>0}$ is equivalent to $(
F_n)_{n>0}$ and therefore belongs to $K^*$. Choose $N>0$ such that ${\mathcal{N}}_{\frac{1}{N}}(x)\subseteq V$.
Define ${\mathcal{I}}_K=[( E_n)_{n>0}]$, where
$E_n=F_N'+\frac{1}{n}$. For any $n>0$, $E_n
\prec F'_N$. Thus $K^* \in \Uun {\mathcal{I}}_K$. We prove
\[
\hspace{5.3cm}\Uun {\mathcal I}_K\cap \phi({\mathcal K}_0(X))\subseteq
\phi(\diamondsuit V).\hspace{5cm}(**)
\]
Assume that ${\mathcal I}_K\ll B^*$. There is $H\in B^*$ such that for
any $n>0$, $E_n\prec H\prec {B}$. Hence, one can
find $(b,0)\in {B}$ with $(x,\frac{1}{N})\prec (b,0)$. Thus
$b\in {\mathcal{N}}_{\frac{1}{N}}(x)$ and $b\in V$. That implies
$B^*\in \phi(\diamondsuit V)$.\qed

It is known that for any quasi-metric space $(X,d)$ whose compact
subsets are $d^{-1}$-precompact, the Vietoris topology on ${\mathcal
K}_0(X)$ coincides with the topology of the Hausdorff
quasi-pseudometric $H_d$ (\cite{rodri-rom}, Theorem 5). So in the
light of Theorem \ref{vietoris}, under this assumption, the map
$\phi:({\mathcal K}_0(X),H_d)\rightarrow ({\bf C}{\bf B}X,\sigma)$ is
an embedding. In the following, we present an alternative proof
which avoids this well-known result.

\begin{thm}\label{hausdorff}
Let $(X,d)$ be a sequentially Yoneda-complete $T_1$ quasi-metric
space such that any compact subset of $(X,d)$ is $d^{-1}$-precompact.
Then the map $\phi:({\mathcal K}_0(X),H_d)\rightarrow({\bf C}{\bf
B}X,\sigma)$ is an embedding.
\end{thm}

\proof Let $\Uun {\mathcal I}$ be a basic open set  of ${\bf
C}{\bf B}X$ in the Scott topology. Let $K_0^*\in \Uun {\mathcal I}$ and
$(F_n)_{n>0}$ be a representation of $K_0^*$. By definition of the
way-below relation, there is a natural number $N>0$ such that for
any $G\in {\mathcal I}$, $G \prec F_N$. Put
$\epsilon=\frac{1}{2}\delta(F_N,F_{N+1})$. To complete the proof
of continuity of $\phi$, it suffices to show that
\[
{\mathcal N}_\epsilon(K_0)\subseteq \phi^{-1}(\Uun {\mathcal I}).
\]
Let $K\in {\mathcal N}_\epsilon(K_0)$. Define $F'=F_{N+1}+\epsilon$. By Remark \ref{rem1}(2),  $F_N\prec F'$.

We prove that $F'\prec {K}$. For this, take
$(y,s+\epsilon)\in F'$. Hence $(y,s)\in F_{N+1}$ and by
$F_{N+1}\prec {K_0}$, there is an element $x\in K_0$ such that
$(y,s)\prec (x,0)$. Since $H_d(K_0,K)<\epsilon$, there is an $x^*\in
K$ such that $d(x,x^*)<\epsilon$. Thus
\[
d(y,x^*)\leq d(y,x)+d(x,x^*)<s +\epsilon.
\]
Consequently $(y,s+\epsilon)\prec (x^*,0)$ and  $F'\prec_L
{K}$. A similar argument shows that $F'\prec_U {K}$
and therefore $F'\prec{K}$ is established. Now, by Lemma
\ref{1}, since $F_N\prec F'\prec{K}$, there is $H\in K^*$
such that $F_N\sqsubseteq H$. Thus for any $G\in {\mathcal I}$, $G \prec H$
and ${\mathcal I}\ll K^*$, as required.

Next, in order to show that the map $\phi$ is an embedding, we
prove that $\phi({\mathcal{N}}_\epsilon(K))$ is a relative Scott
open, for any basic open set ${\mathcal{N}}_\epsilon(K)$. By the
assumption, $K$ is $d^{-1}$-precompact. Hence there is a finite
subset $\{k_1,\ldots,k_m\}$ of $K$ such that for any $k\in K$,
$d(k,k_i)<\epsilon/4$, for some $1\leq i\leq m$. Assume that $(
F_n)_{n>0}$ is a standard representation of $K^*$. Choose
sufficiently large $N>0$ such that $rF_N<\epsilon/4$ and at the
same time for any $k_i\in \{k_1,\ldots,k_m\}$ there is
$(x_i,r_i)\in F_N$ so that $d^*(k_i,x_i)<\epsilon/4$. The latter
property can be achieved by Lemma \ref{way}. Set ${\mathcal I}=[(
E_n)_{n>0}]$, where $E_n=F_N+\frac{1}{n}$. Clearly $K^*\in
\Uun{\mathcal I}$. Now, for proving
\[
\Uun {\mathcal I}\cap \phi({\mathcal K}_0(X))\subseteq
\phi({\mathcal{N}}_\epsilon(K)),
\]
take $B^*\in\Uun{\mathcal I}$ and show that $H_d(K,B)<\epsilon$. Let
$k\in K$. Select $k_i$, $(x_i,r_i)\in F_N$ and (similar to the proof
of $(**)$ in the preceding theorem) $(b,0)\in{B}$ such that
$d(k,k_i)<\epsilon/4$, $d(k_i,x_i)<\epsilon/4$ and
$(x_i,r_i)\sqsubseteq(b,0)$. Therefore
\[
\begin{array}{ll}
d(k,b)&\leq d(k,k_i)+d(k_i,x_i)+d(x_i,b)\\
&<\frac{\epsilon}{4}+\frac{\epsilon}{4}+\frac{\epsilon}{4}=\frac{3\epsilon}{4}.
\end{array}
\]
In other words, $H^-_d(K,B)<\epsilon$. Next, for
$H^+_d(K,B)<\epsilon$, we pick up $b\in B$. One can use the same
argument used in the proof of $(*)$ in the preceding theorem to
find $(x,r)\in F_N$ such that $d(x,b)\leq r<\epsilon/3$. Since
$x\in K$, $H^+_d(K,B)<\epsilon$
follows.\qed

Roughly speaking, the above theorems state that under certain
conditions on $(X,d)$ the hyperspace ${\mathcal K}_0(X)$ with respect
to the Vietoris or Hausdorff topologies can be embedded in a
suitable continuous $\omega$-dcpo. Hence, in the light of
Definition \ref{computational}, the following theorem is obtained.

\begin{thm}\label{wdomrep}
Let $(X,d)$ be a sequentially Yoneda-complete $T_1$ quasi-metric
space. Then
\begin{enumerate}[\em(1)]
\item  The pair  $({\bf C}{\bf B}X,\phi)$
gives an $\omega$-computational model for $({\mathcal
K}_0(X),\tau_V)$.
\item If, in addition, any compact subset of $(X,d)$
is $d^{-1}$-precompact, then the pair  $({\bf C}{\bf B}X,\phi)$
gives an $\omega$-computational model for $({\mathcal K}_0(X),H_d)$.\qed
\end{enumerate}
\end{thm}

\noindent Also, in the light of  Fact \ref{idealcompletion} and Theorem
\ref{dcpo}, since the way-below relation of any continuous poset
is approximating, the ideal completion of $({\bf C}{\bf B}X,\sqsubseteq)$
with the auxiliary relation $\ll$ gives a computational model of
${\mathcal K}_0(X)$. So the following theorem is established.

\begin{thm}\label{domrep}
Let $(X,d)$ be a sequentially Yoneda-complete $T_1$ quasi-metric
space. Then
\begin{enumerate}[\em(1)]
\item  $({\mathcal K}_0(X),\tau_V)$ has a computational model.
\item If, in addition, any compact subset of $(X,d)$
is $d^{-1}$-precompact, $({\mathcal K}_0(X),H_d)$ has a computational
model.\qed
\end{enumerate}
\end{thm}\medskip

\noindent The following corollary is a trivial consequence of the above
Theorems. See also \cite{edalat,L-K,martin1}.

\begin{cor}
Let $(X,d)$ be a complete metric space. Then the hyperspace $({\mathcal
K}_0(X),H_d)$ has a computational model.\qed
\end{cor}

It could be readily seen that the Vietoris and respectively
Hausdorff topologies are $T_1$  if and only if the $\phi$-image of
${\mathcal K}_0(X)$ is a subset of the maximal elements of ${\bf C}{\bf
B}X$. It is known that both these topologies are not necessarily
$T_1$. Therefore the $\phi$-image of ${\mathcal K}_0(X)$ may not lie
in the maximal elements of ${\bf C}{\bf B}X$. As it was noted
before, if $(X,d)$ is Hausdorff (more generally KC), these
topologies are $T_1$.

The following examples give an application of Theorem \ref{domrep}.
\begin{exa}\label{sor}\hfill
\begin{enumerate}[(1)]

\item  Let ${\mathbb R}$ be the set of real numbers and let $d$ be a
$T_1$ quasi-metric defined on $\mathbb{R}$ by $d(x,y)= y-x$ if $x
\leq y$ and $d(x,y)= 1$ if $x > y$. Then the topology $\tau_d$ is
the Sorgenfrey topology on $\mathbb R$ and
${\mathbb{R}}_l=(\mathbb{R},d)$ called the  Sorgenfrey line. It is
easy to see  that $(\mathbb{R},d)$ is a Hausdorff (sequentially)
Yoneda-complete space. It is a well-known fact that any compact
subset of Sorgenfrey line is compact with respect to the usual
topology of ${\mathbb R}$ and hence it is $d^{-1}$-precompact.

\item  Let $\Sigma$ be a non-empty set and $\Sigma^{\infty}$ be
the  set of finite and infinite sequences over $\Sigma$. Define
the relation $\preceq$ on $\Sigma^{\infty}$ as $$x\preceq
y\;\;\;\;\  \leftrightarrow \;\; x\ \mbox{is a prefix of}\ y.$$
For $x,y\in\Sigma^{\infty}$, we denote the longest common prefix
of $x$ and $y$ by  $x\sqcap y$. Also the length of an element
$x\in \Sigma^{\infty}$, is denoted by $l(x)\in \mathbb{N}\cup
\{\infty\}$. Define $q_b: \Sigma^{\infty}\times
\Sigma^{\infty}\rightarrow [0,1]$, given as:

$$q_b(x,y)=2^{-l(x)}-2^{-l(y)}\;\;\ \mbox{if}\  x\preceq y,$$
$$ q_{b}(x,y)=1\;\;\;\ \mbox{otherwise}.$$

We adopt the convention $\frac{1}{\infty}=0$.

It is proved in \cite{ rodri-rom-val } that $(\Sigma^{\infty},
q_b)$ is Hausdorff and  $(q_b)^{-1}$-right $K$-sequentially
complete. Furthermore, it is quite straightforward to complete
the argument in \cite{ rodri-rom-val } to prove this space is in
fact sequentially Yoneda-complete.

\item  Let $d$ be the restriction of the Sorgenfrey metric defined
in Example 1, to $[0,1]$. Put $X=[0,1]^{[0,1]}$ as the set of all
continuous functions $f:[0,1]\rightarrow [0,1]$ with respect to
the quasi-metric space $([0,1],d)$. For $f,g\in X$, let
$$D(f,g)=\sup_{x\in [0,1]} d(f(x),g(x)).$$ Note that $D(f,g)<1$
forces $f\leq g$. Hence any Cauchy sequence $(f_n)_{n>0}$ should
be eventually increasing. Now, for such a sequence take
$f=\sup_{n\geq n_0} f_n$, where $n_0$ is the index from which the
sequence is increasing. Then, $f$ is the Yoneda-limit of
$(f_n)_{n>0}$ and hence $(X,D)$ is a $T_1$ sequentially
Yoneda-complete space. In fact, it can be readily seen that
$(X,D)$ is a Hausdorff space.

\item   Let ${\mathcal C}=[0,1]^{\omega}$. For $p\in [1,\infty)$
the function $q_p:{\mathcal C}\rightarrow [0,1]$ is defined by
\[
q_p(f)=\big( \sum_{n=0}^{\infty} (2^{-n}f(n))^p\big)^{\frac{1}{p}}.
\]
Then using an argument in \cite{rom-san-val}, Theorem 1, one can
show that the function $f_{q_p}:{\mathcal C}\times {\mathcal C}\rightarrow
[0,1]$ given by
\[
f_{q_p}(f,g)=q_p(f-g)\;\;\;\;\ \mbox{if}\ f\geq g,
\]
\[
f_{q_p}(f,g)=1\;\;\;\;\;\;\;\ \mbox{otherwise},
\]
defines a Hausdorff quasi-metric on $\mathcal C$. Furthermore, by
adopting the proofs of Theorems 3 and 4 in \cite{rom-san-val}, one
can also show that $({\mathcal C},f_{q_p})$ is sequentially
Yoneda-complete. (This quasi-metric is in fact the conjugate of
the quasi-metric $e_{d_p}$ given in \cite{rom-san-val}.)
\end{enumerate}
\end{exa}

\begin{cor}
Let $(X,d)$ be one of the examples given in \ref{sor}. Then the
space $({\mathcal K}_0(X),\tau_V)$ is $T_1$ and has a computational
model. Furthermore in the case of Sorgenfrey line $\mathbb{R}_l$,
the space  $({\mathcal K}_0(\mathbb{R}_l),H_d)$ is $T_1$ and has a
computational model.\qed
\end{cor}

\section{A quantitative $\omega$-computational model of ${\mathcal K}_0(X)$}

So far we have shown that, under certain circumstances, $({\mathcal
K}_0(X),H_d)$ can be embedded in ${\bf C}{\bf B}X$ and therefore
$({\bf C}{\bf B}X,\phi)$ provides an $\omega$-computational model
for $({\mathcal K}_0(X),H_d)$. In this section, we take the well-known
quantitative ($\omega$-)computational model approach, proving
that ${\bf C}{\bf B}X$ is also a quantitative
$\omega$-computational model for $({\mathcal K}_0(X),H_d)$ and so the
results of the preceding section are strengthened. Our definition
of quantitative ($\omega$-)computational model follows Rutten
(\cite{Rutten}, Section 7).

\begin{defi}\label{quantitative}
A {\em quantitative ($\omega$-)computational model} of a quasi-metric
space $(Y,d)$ is a quadruple $( L,\sqsubseteq, D, \phi )$ where
$(L,\sqsubseteq)$ is a continuous ($\omega$-)dcpo, $D$ is an algebraic sequentially
Yoneda-complete quasi-metric on $L$ and $\phi:Y\rightarrow L$ is a
map such that:
\begin{enumerate}[(1)]
\item The specialization partial order $\sqsubseteq_D$ is equivalent to the partial order of $L$.
\item $\phi$ is an isometry from $(Y,d)$ into $(L,D)$.
\item $\phi(Max(Y,\sqsubseteq_d))=Max(L,\sqsubseteq)$.
\end{enumerate}
\end{defi}

It is worth mentioning that the above definition is weaker than the definition of Romaguera and Valero (Definition 1 in
\cite{Rom-Val}) in which $(L,D)$ is considered to be
Smyth-complete and the topology $\tau_D$ coincides with the Scott
topology $\sigma_L$. We prefer to take Rutten's notion of
quantitative $\omega$-computational model and then study a special
case where $(L,D)$ satisfies the extra conditions of Definition 1
in \cite{Rom-Val} (Theorems \ref{smyth}).

In \cite{Rom-Val}, Romaguera and Valero followed the work of
Heckmann \cite{Heckmann} for a complete weighted quasi-metric
space $(X,d)$ and defined a complete partial quasi-metric $Q$ on
the space of formal balls. A quasi-metric $q$ on $\mathbf{B}X$ is
then derived from $Q$ which induces the same topology as $Q$ on
$\mathbf{B}X$ and moreover, $(\mathbf{B}X,q)$ is Smyth-complete.
It is useful to note that the quasi-metric $q$ can be defined
directly on $\mathbf{B}X$ for any quasi-metric space without the
existence of a partial metric.

\begin{defi}\label{q}
Let $(X,d)$ be a quasi-metric space. For $(x,r),(y,s)\in
\mathbf{B}X$, define
\[
q((x,r),(y,s))=\max\{d(x,y),|r-s|\}+(s-r).
\]
\end{defi}

It is easy to see that $q$ defines a quasi-metric on $\mathbf{B}X$.
The next lemma shows that $(\mathbf{B}X,q)$ inherits
Smyth-completeness and sequentially Yoneda-completeness from
$(X,d)$. The proof is more or less the same as the proof of Theorem 4.1 of \cite{Rom-Val}.

\begin{lem}\label{Smyth2}
Let $(X,d)$ be a sequentially Yoneda-complete (respectively
Smyth-complete) quasi-metric space. Then $(\mathbf{B}X,q)$ is also
sequentially Yoneda-complete (respectively Smyth-complete).\qed
\end{lem}

The following theorem generalizes Theorem 5.1 of \cite{Rom-Val}.

\begin{thm}
Each algebraic sequentially Yoneda-complete $T_1$ quasi-metric space
has a quantitative computational model.
\end{thm}

\proof Let $(X,d)$ be an algebraic sequentially
Yoneda-complete $T_1$ quasi-metric space and $q$ be the
quasi-metric defined in Definition \ref{q} on $\mathbf{B}X$. By
Theorem \ref{2}, $\mathbf{B}X$ is a continuous dcpo. Also, by the
above lemma $(\mathbf{B}X,q)$ is sequentially Yoneda-complete. A
straightforward computation shows that for any finite element
$x\in X$, $(x,r)$  is finite in $(\mathbf{B}X,q)$ and moreover
the set of such elements forms a base for $(\mathbf{B}X,q)$.
Therefore $(\mathbf{B}X,q)$ is algebraic. The other parts follow
easily from the proof of Theorem
4.1 of \cite{Rom-Val}.\qed

Now, we turn to the main topic of this section.   We wish to define
a quasi-metric $D$ on ${\bf C}{\bf B}X$ and show that $({\bf C}{\bf
B}X,D)$ together with the map $\phi:{\mathcal K}_0(X)\rightarrow {\bf
C}{\bf B}X$ which is defined as $\phi(K)=K^*$, form a quantitative
computational model for $({\mathcal K}_0(X),H_d)$.

To emphasize, we fix $(X,d)$ to be a sequentially Yoneda-complete
$T_1$ quasi-metric space, though $(X,d)$ need not be algebraic.

\begin{defi}
Let $q$ be the quasi-metric defined in Definition \ref{q} on
${\bf B}X$. Recall that $H_q$ on ${\mathcal P}_{fin}{\bf B}X$ is defined
by
\[
H_q(F,G)=\max\{\sup_{(x,r)\in F}\inf_{(y,s)\in
G}q((x,r),(y,s)),\sup_{(y,s)\in G}\inf_{(x,r)\in F}q((x,r),(y,s))\},
\]
for any $F,G\in {\mathcal P}_{fin}{\bf B}X$. Put $D$ on ${\bf C}{\bf
B}X$ as follows:
\[
\ \ D({\mathcal I},{\mathcal J})=\sup_{F_n}\inf_{G_m}H_d(F_n,G_m),
\]
for all ${\mathcal I},{\mathcal J}\in {\bf C}{\bf B}X$ and representations
$(F_n)_{n>0},( G_m)_{m>0}$ of ${\mathcal I}$
and ${\mathcal J}$, respectively.
\end{defi}

Next, we  show that $D$ is independent from any particular choice of
representations.
\begin{lem}\label{well-defined2}
$D$ is well-defined on ${\bf C}{\bf B}X$.
\end{lem}
\proof Let $(F_n)_{n>0}$ and $(F'_k)_{k>0}$ be two different representations for ${\mathcal I}$. Since
$(F_n)_{n>0} \sim (F'_k)_{k>0}$, for any
$F_n$ there is $F'_k$ such that $F_n\prec F'_k$. So
$H_q(F_n,F'_k)=0$ and therefore
\[
H_q(F_n,G_m)\leq H_q(F'_k,G_m).
\]
So by taking infimum on $G_m$, we have
\[
\inf_{G_m}H_q(F_n,G_m)\leq\inf_{G_m}H_q(F'_k,G_m).
\]
Thus
\[
\inf_{G_m}H_q(F_n,G_m)\leq\sup_{F'_k}\inf_{G_m}H_q(F'_k,G_m),
\]
\[
\sup_{F_n}\inf_{G_m}H_q(F_n,G_m)\leq\sup_{F'_k}\inf_{G_m}H_q(F'_k,G_m).
\]
Similarly, we can prove that
\[
\sup_{F'_k}\inf_{G_m}H_q(F'_k,G_m)\leq\sup_{F_n}\inf_{G_m}H_q(F_n,G_m).\eqno{\qEd}
\]

\begin{prop}\label{order}
$D$ is a quasi-metric on ${\bf C}{\bf B}X$. In addition, the
specialization order $\sqsubseteq_D$ is equivalent to the partial
order $\sqsubseteq$ defined on ${\bf C}{\bf B}X$.
\end{prop}
\proof The triangular inequality is straightforward. So we
only check that $D({\mathcal I},{\mathcal J})=D({\mathcal J},{\mathcal I})=0$
implies ${\mathcal I}={\mathcal J}$. Let $D({\mathcal I},{\mathcal J})=0$ and $(
F_n)_{n>0}$ and $( G_m)_{m>0}$ be representations for ${\mathcal I}$
and ${\mathcal J}$, respectively. Fix $n>0$ and put
$\delta=\delta(F_{n},F_{n+1})$. From the definition of $D$, there
is $m>0$ such that $H_q(F_{n+1},G_m)<\delta$. We prove $F_n\prec
G_m$. Let $(x,r)\in F_n$. There is $(x',r')\in F_{n+1}$ with
$(x,r)\prec(x',r')$. Also, by $H_q(F_{n+1},G_m)<\delta$, there is
$(y,s)\in G_m$ such that $q((x',r'),(y,s))<\delta$. This means
that $d(x',y)<r'-s+\delta$ and consequently
\[
\begin{array}{ll}
d(x,y)&\leq d(x,x')+d(x',y)\\
&<(r-r')-\delta+(r'-s)+\delta\\
&=r-s.
\end{array}\]
Thus $(x,r)\prec(y,s)$ and therefore $F\prec_L G$. $F\prec_U G$ can
be shown in a similar fashion. Hence  $F\prec G$ is established and
therefore ${\mathcal I}\sqsubseteq{\mathcal J}$. Now $D({\mathcal J},{\mathcal
I})=0$ also implies that ${\mathcal J}\sqsubseteq{\mathcal I}$.

For the second part, clearly by the above argument, $D({\mathcal
I},{\mathcal J})=0$ implies ${\mathcal I}\sqsubseteq{\mathcal J}$. Conversely,
let ${\mathcal I}\sqsubseteq{\mathcal J}$. So for any $F\in{\mathcal I}$, there
is $G\in {\mathcal J}$  with $F\prec G$. Thus $H_q(F,G)=0$ which implies
$D({\mathcal I},{\mathcal J})=0$.\qed

\begin{prop}\label{Yoneda}
The domain ${\bf C}{\bf B}X$ equipped with the quasi-metric $D$ is
sequentially Yoneda-complete.
\end{prop}
\proof Let $({\mathcal I}_n)_{n>0}$ be a Cauchy sequence in
$({\bf C}{\bf B}X,D)$ and for any $n>0$, $( F^n_m)_{m>0}$ be a
representation for ${\mathcal I}_n$. For any $n>0$, there is a
natural number $N_n>n$ such that $N_n > N_{n-1}$ for $n > 1$ and for any $l\geq k\geq N_n$,
$D({\mathcal I}_k,{\mathcal I}_l)<\frac{1}{2^{n+1}}$, i.e.
\[
\forall F^k_r\in {\mathcal I}_k\ \ \exists F^l_s\in{\mathcal I}_l \ \
H_q(F^k_r,F^l_s)<\frac{1}{2^{n+1}}.
\]
Define $( G_{ij})_{i,j>0}$ as follows:\\
Fix $F^{N_1}_1\in{\mathcal I}_{N_1}$ and put $G_{11}=F^{N_1}_1$.
Inductively for any $k\geq 2$, choose $G_{1k}\in (F^{N_k}_m)_{m>0}$ such that
$$H_q(G_{1{k-1}},G_{1k})<\frac{1}{2^k}.$$
In a similar way, for any $i\geq 2$, put $G_{i1}=F^{N_1}_i$ and
inductively  find $(G_{ik})_{k>0}$ such that for any
$i<j$, $G_{ik}\prec G_{jk}$ and
\[
H_q(G_{i{k-1}},G_{ik})<\frac{1}{2^k}.
\]
Let $L_k=G_{kk}+\frac{1}{2^{k-1}}$. The
following two claims complete the proof.\\

{\bf Claim 1.} ${\mathcal I}=[( L_k)_{k>0}]\in {\bf C}{\bf
B}X$, i.e. $( L_k)_{k>0}$ is an ascending sequence in
${\mathcal P}_{fin}{\bf B}X$.\\
Let $k>0$ and $(x,r+\frac{1}{2^{k-1}})\in L_k$. Hence $(x,r)\in
G_{kk}$. Since
\[
\begin{array}{ll}
H_q(G_{kk},G_{k+1k+1})&\leq H_q(G_{kk},G_{k+1k})+H_q(G_{k+1k},G_{k+1k+1})\\
&<0+\frac{1}{2^{k+1}}
\end{array}
\]
there is a $(y,s)\in G_{k+1k+1}$ such that $q((x,r),(y,s))<\frac{1}{2^{k+1}}$. Hence
$d(x,y)<r-s+\frac{1}{2^{k+1}}$. By a simple calculation,
\[
d(x,y)<(r+\frac{1}{2^{k-1}})-(s+\frac{1}{2^{k}}).
\]
Therefore $(x,r+\frac{1}{2^{k-1}})\prec(y,s+\frac{1}{2^{k}})$,
where $(y,s+\frac{1}{2^{k}})\in L_{k+1}$. This means that
$L_k\prec_LL_{k+1}$. Similarly, we can prove  $L_k\prec_UL_{k+1}$ and
consequently $L_k\prec L_{k+1}$.\\

{\bf Claim 2.} $\mathcal I$ is the Yoneda limit of $({\mathcal
I}_n)_{n>0}$. \\
First, we prove that $D({\mathcal I}_n,{\mathcal I})\rightarrow 0$. Let
$\epsilon>0$. Since  $({\mathcal I}_n)_{n>0}$ is a Cauchy sequence,
there is $N>0$ such that for any $r\geq s\geq N$, $D({\mathcal
I}_s,{\mathcal I}_r)<\frac{\epsilon}{3}$. Fix $n>N$ and $F\in {\mathcal
I}_n=[( F^n_m)_{m>0}]$. Choose $N_k>n$ from the cofinal sequence
$(N_n)_{n>0}$  constructed above  such that
$\frac{1}{2^{k-1}}<\frac{\epsilon}{3}$. Since $D({\mathcal I}_n,{\mathcal
I}_{N_k})<\frac{\epsilon}{3}$, there is $F^{N_k}_{m_0}\in {\mathcal
I}_{N_k}$ such that $H_q(F,F^{N_k}_{m_0})<\frac{\epsilon}{3}$.
Choose $G_{lk}\in {\mathcal I}_{N_k}$ with $F^{N_k}_{m_0} \prec
G_{lk}$ and $l>k$. Now,
$H_q(G_{lk},G_{ll})<\frac{1}{2^{k+1}}+\cdots+\frac{1}{2^l}<\frac{1}{2^k}<\frac{\epsilon}{3}$ implies
that
\[
\begin{array}{ll}
H_q(F,L_l)&\leq H_q(F,F^{N_k}_{m_0})+H_q(F^{N_k}_{m_0},G_{lk})+
H_q(G_{lk},G_{ll})+H_q(G_{ll},L_l)\\&<
\frac{\epsilon}{3}+0+\frac{\epsilon}{3}+\frac{\epsilon}{3}\\&=\epsilon.
\end{array}
\]
Now, we show that $\mathcal I$ is the Yoneda limit of $({\mathcal
I}_n)_{n>0}$. Because of $D({\mathcal I}_n,{\mathcal I})\rightarrow 0$,
it suffices to prove that $D({\mathcal I},{\mathcal J})\leq
\inf_n\sup_{k>n} D({\mathcal I}_k,{\mathcal J})$, for any ${\mathcal J}\in
\mathbf{C}\mathbf{B}X$. Put $\inf_n\sup_{k>n} D({\mathcal I}_k,{\mathcal
J})=s$. We prove that for any $\epsilon>0$, $D({\mathcal I},{\mathcal
J})\leq s+\epsilon$. Let $L\in {\mathcal I}$. Since $\inf_n\sup_{k>n}
D({\mathcal I}_k,{\mathcal J})=s$, there is $N>0$ such that for any
$n>N$, $D({\mathcal I}_n,{\mathcal J})\leq s+\epsilon/2$. Take $k>0$ such
that $L\prec L_k$, $\frac{1}{2^{k-1}}<\epsilon/2$ and $N_k>N$.
So, for any representation $(G_m)_{m>0}$ of $\mathcal J$, there is
$m>0$ such that
\[
\hspace{2.8cm}\begin{array}{ll} H_q(L,G_m)&\leq
H_q(L,L_k)+H_q(L_k,G_{kk})+H_q(G_{kk},G_m)\\&<0+\frac{1}{2^{k-1}}+
s+\epsilon/2\\&<s+\epsilon.
\end{array}
\]
This complete the proof.\qed

Below, we show that the quasi-metric space $({\bf C}{\bf B}X,D)$
is in fact the sequential Yoneda completion of $({\mathcal
P}_{fin}{\mathbf B}X,H_d)$. First, we recall the definition of
sequential Yoneda completion \cite{kunzi-schell}.

\begin{defi}\label{ycompletion}
Let $(Y,d)$ be a quasi-metric space and
\[
\widehat{Y}=\{(x_n)_{n>0}\ :\ (x_n)_{n>0}\
\mbox{is a Cauchy sequence}\}.
\]
Define the quasi-pseudometric
$\widehat{d}$ and the equivalence relation $\approx$ as follows:
\[
\widehat{d}((x_n)_{n>0},(y_m)_{m>0})=\inf_n\sup_{k\geq n}\sup_m\inf_{p\geq m}d(x_k,y_p),
\]
\[
(x_n)_{n>0}\approx(y_m)_{m>0}\
\Leftrightarrow\ \widehat{d}((x_n)_{n>0},(y_m)_{m>0}) =\widehat{d}((y_m)_{m>0},(x_n)_{n>0})=0.
\]
Put
$\overline{Y}=\widehat{Y}/\approx$ and define the quasi-metric $\overline{d}$ by
\[
\overline{d}([(x_n)_{n>0}],[(y_m)_{m>0}])=\widehat{d}((x_n)_{n>0},(y_m)_{m>0}).
\]
The pair $(\overline{Y},\overline{d})$ is called the sequential
Yoneda completion of $(Y,d)$.
\end{defi}

\begin{prop}\label{completion}
$({\bf C}{\bf B}X,D)$ is the sequential Yoneda completion of
$({\mathcal P}_{fin}{\mathbf B}X,H_d)$.
\end{prop}
\proof Let $(F_n)_{n>0}$ be a Cauchy sequence
in ${\mathcal P}_{fin}{\mathbf B}X$ and $\overline{H}_d$ be the
quasi-metric completion of $H_d$ (Definition \ref{ycompletion}).
Without loss of generality, we can assume that for any $0<n\leq m$,
$H_d(F_n,F_m)<\frac{1}{2^{n+1}}$, since there is an
$\overline{H}_d$-equivalent subsequence of $(F_n)_{n>0}$,
satisfying this property. Now, we show that there is an ascending
chain $(L_k)_{k>0}$ in ${\mathcal P}_{fin}{\mathbf B}X$ which
is $\overline{H}_d$-equivalent. For any $n>0$, define
${\mathcal I}_n=[( E^n_m)_{m>0}]$ where
$E^n_m=F_n+\frac{1}{m}$. One can readily check
that $({\mathcal I}_n)_{n>0}$ is a Cauchy sequence. For $k>0$, put $L_k=E^k_k+\frac{1}{2^{k-1}}$.
A similar argument as used in Claim 1 of Proposition \ref{Yoneda},
shows that $[(L_k)_{k>0}]$ is a $\prec$-ascending chain. In order to show
\[
\overline{H}_d((F_n)_{n>0},(L_m)_{m>0})=0,
\]
we have to verify the following:
\[
\forall \epsilon>0 \ \exists n\ \forall k\geq n\ \forall m\ \exists p\geq m \ H_d(F_k,L_p)<\epsilon.
\]
Let $\epsilon >0$ be given. Find $n>0$ so that
$\frac{1}{2^{n+1}}<\frac{\epsilon}{2}$. Next, for any $k\geq n$ and
$m>0$, take $p\geq k,m$ such that
$\frac{1}{p}+\frac{1}{2^{p-1}}<\frac{\epsilon}{2}$. Since
$L_p=F_p+ (\frac{1}{p}+\frac{1}{2^{p-1}})$, it
easily follows that $H_d(F_k,L_p)<\epsilon$. The equality
$\overline{H}_d((L_m)_{m>0},(F_n)_{n>0})=0$
can be proved in a similar way.

So, the map 
\[
\begin{array}{llllll} \psi:&(\overline{{\mathcal
P}_{fin}{\mathbf
B}X},\overline{H}_d)&&\longrightarrow&&({\bf C}{\bf B}X,D)\\
&[(F_n)_{n>0}]&&\longmapsto&&[(L_k)_{k>0}].
\end{array}
\]
defines a bijective isometry, as required.\qed

\begin{cor}\label{algebraic}
The domain ${\bf C}{\bf B}X$ equipped with the quasi-metric $D$ is
algebraic.
\end{cor}
\proof By Corollary 23 of \cite{kunzi-schell}, any Yoneda
completion is algebraic. Hence, according to Proposition
\ref{completion}, $({\bf C}{\bf B}X,D)$ is algebraic.\qed

\begin{prop}\label{isometry}
The map $\phi$ is an isometry between $({\mathcal K}_0(X),H_d)$ and
$({\bf C}{\bf B}X,D)$.
\end{prop}
\proof Let $K_1,K_2\in {\mathcal K}_0(X)$ and $( F_n)_{n>0}$
and $(G_m)_{m>0}$ be standard representations for $K_1$ and
$K_2$, respectively.

Assume that $H_d(K_1,K_2)=d$ and $\epsilon>0$. We show that
$D(K^*_1,K^*_2)\leq d+\epsilon$. Fix $n>0$ and let
$\delta=\delta(F_n, F_{n+1})$. We wish to find $m>0$ such that
\[
H_q(F_n,G_M)<d+\epsilon.
\]
Let $(x,r)\in F_n$. Since $F_n\prec F_{n+1}\prec {K_1}$, there are
$(x',r')\in F_{n+1}$ and $(z,0)\in {K_1}$ such that
$d(x,x')<(r-r')-\delta$ and $d(x',z)<r'$. Because of $H_d(K_1,K_2)=d$,
there is $y\in K_2$ such that $d(z,y)\leq d+\epsilon$. By Lemma
\ref{way}, there is a sequence $( (y_m,s_m))_{m>0}$ such
that $(y_m,s_m)\in G_m$ and $d^*(y_m,y)\rightarrow 0$. Select $y_m$
such that $d(y,y_m)<\frac{\delta}{2}$ and $s_m<\frac{\delta}{2}$.
Then
\[
\begin{array}{ll}
d(x,y_m)&\leq d(x,x')+d(x',z)+d(z,y)+d(y,y_m)\\
 &<(r-r')-\delta+r'+d+\epsilon+\frac{\delta}{2}\\
 &=r+d+\epsilon-\frac{\delta}{2}\\
\end{array}
\]
Thus for any $(x,r)\in F_n$, there is $(y_{m_x},s_{m_x})\in G_{m_x}$
such that $d(x,y_{m_x})<r-s_{m_x}+d+\epsilon$. Take
$M\geq\max\{m_x:\ (x,r)\in F_n\}$ with $rG_M<\frac{\delta}{2}$.
Hence, by the above calculation, for any $(x,r)\in F_n$ there exists
$(y,s)\in G_M$ such that $q((x,r),(y,s))<d+\epsilon$. Similarly, for
any $(y,s)\in G_M$ one can find $(x,r)\in F_n$ with
\[
q((x,r),(y,s))<d+\epsilon.
\]
So $H_q(F_n,G_M)<d+\epsilon$ and therefore $D(K^*_1,K^*_2)\leq
H_d(K_1,K_2)$.

On the other hand, we show that $H_d(K_1,K_2)\leq D(K^*_1,K^*_2)$.
Let $D(K^*_1,K^*_2)=d$ and $x\in K_1$. Again, Lemma \ref{way}
implies the existence of a sequence $( (x_n,r_n))_{n>0}$
such that $(x_n,r_n)\in F_n$ and $d^*(x_n,x)\rightarrow 0$. For $\epsilon>0$, choose
$(x_N,r_N)$ such that $d(x,x_N)<\frac{\epsilon}{3}$ and $r_N<\frac{\epsilon}{3}$.
Since $D(K^*_1,K^*_2)=d$,  there is $G_{m_N}$ and
$(y_{m_N},s_{m_N})\in G_{m_N}$ such that
$d(x_N,y_{m_N})<r_N-s_{m_N}+d+\frac{\epsilon}{3}$. The following
inequalities
\[
\begin{array}{ll}
d(x,y_{m_N})&\leq d(x,x_N)+d(x_N,y_{m_N})\\&<
\frac{\epsilon}{3}+r_N-s_{m_N}+d+\frac{\epsilon}{3}\\&< d+\epsilon,
\end{array}
\]
imply $\inf_{y\in K_2}d(x,y)\leq d$ and therefore
\[
\sup_{x\in K_1}\inf_{y\in K_2}d(x,y)\leq d.
\]
Now, in a similar way, $\sup_{y\in K_2}\inf_{x\in K_1}d(x,y)\leq d$.
Hence $H_d(K_1,K_2)\leq d$ and consequently $H_d(K_1,K_2)\leq
D(K^*_1,K^*_2)$.\qed

Now, the main theorem of this section is stated.

\begin{thm}
$({\bf C}{\bf B}X,D)$ together with the map $\phi:{\mathcal
K}_0(X)\rightarrow{\bf C}{\bf B}X $ form a quantitative
$\omega$-computational model for $({\mathcal K}_0(X),H_d)$.
\end{thm}
\proof By the above propositions, $({\bf C}{\bf B}X,D)$ is an
algebraic sequentially Yoneda-complete quasi-metric space, the
specialization partial order $\sqsubseteq_D$ is equivalent to the
partial order of ${\bf C}{\bf B}X$ and the map $\phi$ is an isometry
between $({\mathcal K}_0(X),H_d)$ and $({\bf C}{\bf B}X,D)$. The
condition $\phi(Max({\mathcal K}_0(X),\sqsubseteq_{H_d})) =Max({\bf
C}{\bf B}X,\sqsubseteq)$ follows from propositions \ref{maximal} and
\ref{order}.\qed

Next, we impose some extra conditions on $(X,d)$ under which this space  has
a quantitative $\omega$-computational model in the sense of Romaguera and
Valero \cite{Rom-Val}. We, hereby, suppose that $(X,d)$ is
Smyth-complete and any compact subset of $(X,d)$ is
$d^{-1}$-precompact. These conditions guarantee that any Cauchy
sequence in $({\mathcal P}_{fin}{\bf B}X,H_q)$ is biCauchy.

\begin{lem}\label{biCauchy}
Let $(X,d)$ be a Smyth-complete quasi-metric space of which all
of its compact subsets are $d^{-1}$-precompact. Then any
$\omega$-chain in ${\mathcal P}_{fin}{\bf B}X$ is  biCauchy.
\end{lem}
\proof First note, one can easily show that any compact
subset of $({\bf B}X,q)$ is $q^{-1}$-precompact. Now, let $(
F_n)_{n>0}$ be an $\omega$-chain in ${\mathcal P}_{fin}{\bf B}X$. For
any $n<m$, $F_n\prec F_m$. Hence $H_q(F_n,F_m)=0$ and $(
F_n)_{n>0}$ is a Cauchy sequence. According to Lemma
\ref{Smyth2}, $({\bf B}X,q)$ is Smyth-complete. So by Theorem
\ref{smythcomplete}, $({\mathcal K}_0({\bf B}X),H_q)$ is
Smyth-complete. Therefore ${\mathcal P}_{fin}{\bf B}X$ as a subspace
of ${\mathcal K}_0({\bf B}X)$ is Smyth-completable and so any Cauchy
sequence in $({\mathcal P}_{fin}{\bf B}X,H_q)$ is biCauchy. Hence $(
F_n)_{n>0}$ is a biCauchy sequence.\qed

The following auxiliary lemma will be useful in several proofs.
\begin{lem}\label{3}
Let $G\prec H\sqsubseteq M$ and $H_q(M,L)<\delta(G,H)$. Then $G\prec L$.
\end{lem}
\proof Let $(x,r)\in G$ and $\delta=\delta(G,H)$. Then there are $(y,s)\in H$,
$(z,t)\in M$ and $(a,u)\in L$ such that
\[
(x,r)\prec
(y,s)\sqsubseteq (z,t)\;\;\ \mbox{and}\;\;\  q( (z,t),(a,u))<\delta.
\]
Then the following inequalities follow:
\[
\begin{array}{ll}
d(x,a)&\leq d(x,y)+ d(y,z)+d(z,a)\\
&<(r-s)-\delta+(s-t)+(t-u)+\delta\\
&=r-u.
\end{array}
\]
Hence $(x,r)\prec (a,u)$ and consequently $G\prec_L L$.
A similar calculation shows that $G\prec_U L$ and therefore
$G\prec L$ as required.\qed

\begin{thm}\label{smyth}
Under the assumptions of Lemma \ref{biCauchy},
\begin{enumerate}[\em(1)]
\item $({\bf C}{\bf B}X,D)$ is Smyth-complete.
\item The topology $\tau_D$ induced by $D$ coincides with the Scott topology
of the domain ${\bf C}{\bf B}X$.
\end{enumerate}
\end{thm}
\proof (1) Let $({\mathcal I}_n)_{n>0}$ be a Cauchy sequence in
$({\bf C}{\bf B}X,D)$ and for any $n>0$, $( F^n_m)_{m>0}$ be a
representation for ${\mathcal I}_n$. Now, fix the natural sequence
$(N_n)_{n>0}$, the double sequence $(G_{ij})_{i,j>0}$ and ${\mathcal
I}=[( L_k)_{k>0}]$ which are constructed in the proof of
Proposition \ref{Yoneda}. We show that $({\mathcal I}_n)_{n>0}$
converges strongly to $\mathcal I$. Because of $D({\mathcal I}_n,{\mathcal
I})\rightarrow 0$, it suffices to verify $D({\mathcal I},{\mathcal
I}_n)\rightarrow 0$. Let $\epsilon>0$. Since $(L_k)_{k>0}$ is a
Cauchy $\omega$-chain in $({\mathcal P}_{fin}{\bf B}X,H_q)$, so, by
Lemma \ref{biCauchy}, it is a biCauchy sequence. Choose $k>0$
such that $\frac{1}{2^{k-1}}<\epsilon/4$ and for any $r,s\geq k$,
$H_q(L_s,L_r)<\epsilon/4$. We show that for any $n>N_k$, $D({\mathcal
I},{\mathcal I}_n)<\epsilon$.

Take $n>N_k$ and  $L_i\in {\mathcal I}$. Recall that the sequence
$(N_j)_{j>0}$ has the following properties:
\[
\forall r\geq s\geq N_j,\ \ \  D({\mathcal I}_s,{\mathcal
I}_r)<\frac{1}{2^{j+1}}.
\]
Particularly $D({\mathcal I}_{N_k},{\mathcal I}_n)<\frac{1}{2^{k+1}}$. So,
there is $F^n_m\in {\mathcal I}_n$ such that for $G_{kk}\in {\mathcal
I}_{N_k}$, $H_q(G_{kk},F^n_m)<\epsilon/4$. Now, if $i\leq k$,
then $H_q(L_i,L_k)=0$. Otherwise, since $(L_k)_{k>0}$ is a
biCauchy sequence, $H_q(L_i,L_k)<\frac{\epsilon}{4}$. Thus
\[
\begin{array}{ll}
H_q(L_i,F^n_m)&\leq
H_q(L_i,L_k)+H_q(L_k,G_{kk})+H_q(G_{kk},F^n_m)\\&<
\frac{\epsilon}{4}+\frac{1}{2^{k-1}}+\frac{\epsilon}{4}\\&<\frac{3\epsilon}{4}.
\end{array}
\]

(2) Suppose that $\Uun{\mathcal I}$ is a basic open subset of ${\bf
C}{\bf B}X$ and ${\mathcal J}\in\Uun{\mathcal I}$. Let $( G_n)_{n>0}$ be
a representation of $\mathcal J$. So, there is $n>0$ such that
$F\prec G_n$, for any $F\in {\mathcal I}$. Set $\delta=\delta(G_n,
G_{n+1})$. We prove that
\[
{\mathcal N}_\delta(\mathcal J)=\{{\mathcal L}\ :\ \ D({\mathcal J},{\mathcal L})<\delta
\}\subseteq \Uun{\mathcal I}
\]
Let ${\mathcal L}\in {\mathcal N}_\delta(\mathcal J)$ and $(L_m)_{m>0}$ be
one of its representations. From $D({\mathcal J},{\mathcal L})<\delta$, there is $L_m$ such that
$H_q(G_{n+1},L_m)<\delta$. Also, according to the Lemma \ref{3}, $G_n\prec G_{n+1}\sqsubseteq G_{n+1}$ and
$H_q(G_{n+1},L_m)<\delta$ implies that $G_n\prec L_m$. Finally,
since for every $F\in {\mathcal I}$, $F\prec G_n$ it
follows that for any $F\in {\mathcal I}$, $F\prec L_m$. Therefore
${\mathcal L}\in \Uun {\mathcal I}$.

Now, let ${\mathcal N}_\delta(\mathcal I)$ be a basic open subset in
$\tau_D$. Assume that $(F_n)_{n>0}$ is a representation of $\mathcal
I$. According to Lemma \ref{biCauchy}, $(F_n)_{n>0}$ is a biCauchy
sequence. So there is $N>0$ such that
\[
\forall n,m\geq N\ \ \ H_q(F_n,F_m)<\frac{\delta}{2}.
\]
Put ${\mathcal J}=( G_n)_{n>0}$, where
$G_n=F_N+\frac{1}{n}$. Clearly ${\mathcal J}\ll
{\mathcal I}$. We prove that
\[
\Uun{\mathcal J}\subseteq {\mathcal N}_\delta(\mathcal I).
\]
Let ${\mathcal L}\in \Uun{\mathcal J}$ and $( L_m)_{m>0}$ be a
representation for $\mathcal L$. Since ${\mathcal J}\ll {\mathcal L}$, there is
$L_m$ such that for any $n>0$, $G_n\prec L_m$. For any
$F_n\in {\mathcal I}$, if $n<N$ then $H_q(F_n,F_N)=0$. Otherwise
$H_q(F_n,F_N)<\frac{\delta}{2}$. Take $k>0$ with
$\frac{1}{k}<\frac{\delta}{4}$. Then we have
\[
\begin{array}{ll}
H_q(F_n,L_m)&\leq
H_q(F_n,F_N)+H_q(F_N,G_k)+H_q(G_k,L_m)\\&<\frac{\delta}{2}+\frac{1}{k}+0\\&<\frac{3\delta}{4}.
\end{array}
\]
Thus $D({\mathcal I},{\mathcal L})<\delta$ or equivalently ${\mathcal
L}\in {\mathcal N}_\delta({\mathcal I})$.\qed

\begin{rem}
The isometry of $\phi$ (Proposition  \ref{isometry}) and the above
Theorem, imply that for any Smyth-complete $T_1$ quasi-metric space
whose compact subsets are $d^{-1}$-precompact, the map $\phi$ is
an embedding. This result was already proved in Theorem
\ref{hausdorff}.
\end{rem}

\section{Plotkin powerdomain vs. $\omega$-Plotkin domain}
In this section, we compare the Plotkin Powerdomain and
$\omega$-Plokin constructions of $\mathbf{B}X$. The Plotkin
powerdomain of $\mathbf{B}X$, denoted by ${\mathcal P}{\bf B}X$, is
the ideal completion of the abstract basis $({\mathcal
{P}}_{fin}\mathbf{B}X,\prec_{EM})$. In the following, we show
that for any $T_1$ quasi-metric space $(X,d)$ if $(X,d)$ is either
Smyth-complete and all of its compact subsets are
$d^{-1}$-precompact, or $\omega$-algebraic Yoneda-complete,
then ${\mathcal P}{\bf B}X$ and ${\bf C}{\bf B}X$ are
order-isomorphic.

\begin{thm} \label{idch}
Let $(X,d)$ be a $T_1$ quasi-metric space. Assume also that either
of the following conditions hold.
\begin{enumerate}[\em(1)]
\item $(X,d)$ is a Smyth-complete space all of
whose compact subsets are $d^{-1}$-precompact.
\item $(X,d)$ is an $\omega$-algebraic Yoneda-complete space.
\end{enumerate}
Then ${\mathcal P}{\bf B}X$ and ${\bf C}{\bf B}X$ are order-isomorphic.
\end{thm}
\proof (1) Assume first that $(X,d)$ satisfies condition 1
above. Let ${\mathcal I}$ be a round ideal in ${\mathcal P}_{fin}{\bf
B}X$. We claim that  ${\mathcal I}$  has  a cofinal
$\prec_{EM}$-ascending subsequence $(F_n)_{n>0}$. Note that
${\mathcal I}$ can be considered as a Cauchy net in $({\mathcal
P}_{fin}{\bf B}X, H_q)$, since if $G\prec_{EM} H\in {\mathcal I}$,
then $H_q(G,H)=0$. Under the above assumptions, similar to the
argument used in Lemma \ref{biCauchy}, $({\mathcal P}_{fin}{\bf B}X,
H_q)$ is Smyth-completable. Hence ${\mathcal I}$ is biCauchy and has a
biCauchy subsequence $(F_n)_{n>0}$, satisfying the following
property:
\[
\hspace{3.5cm}\forall H,G\in {\mathcal I},\ F_n\prec_{EM} H,G\;\;\ \Rightarrow\;\;\;\  H_q(H,G)<\frac{1}{n}.\hspace{3.1cm}(*)
\]
Without loss of generality, we may assume that $(F_n)_{n>0}$ is
$\prec_{EM}$-ascending. Now, we show that this sequence is
$\prec_{EM}$-cofinal in ${\mathcal I}$. Let $G\in {\mathcal I}$. $N>0$
must be found such that $G\prec_{EM} F_N$. Choose $H\in {\mathcal I}$
with $G\prec_{EM} H$ and put $\delta=\delta(G,H)$. Take $N>1$ such
that $\frac{1}{N-1}<\delta$ and let $M\in {\mathcal I}$ with
$F_N,H\prec_{EM} M$. Therefore by $F_{N-1}\prec_{EM} F_N\prec_{EM} M$ and
$(*)$, we have $H_q(M,F_N)<\frac{1}{N-1}<\delta$. Finally,
according to the Lemma \ref{3}, $G\prec_{EM}H\prec_{EM}M$
and $H_q(M,F_N)<\delta$ implies  $G\prec_{EM} F_N$.

Now, it is easy to check that  the mapping
\[
\begin{array}{llllll} \psi:&{\mathcal P}{\bf B}X&&\longrightarrow&&{\bf C}{\bf B}X\\
& {\mathcal I}&&\longmapsto&&[(F_n)_{n>0}]
\end{array}
\]
defines an order isomorphism.

(2) Now consider that $(X,d)$ is an $\omega$-algebraic
Yoneda-complete space. Let $X_0$ be a countable algebraic  subset
of $X$. Then the set ${\bf B}_{\mathbb{Q}} X_0=X_0\times
\mathbb{Q}^+$ and respectively  the set ${\mathcal P}_{fin}{\bf
B}_{\mathbb{Q}}X_0$ of all finite subsets of ${\bf
B}_{\mathbb{Q}} X_0$ form a countable basis for ${\bf B}X$ and
 for  ${\mathcal P}_{fin}{\bf B}X$ respectively. Now, in the light of
Theorem 6.2.3 of \cite{abramsky}, ${\mathcal P}{\bf B}X$ is given by
the ideal completion of $({\mathcal P}_{fin}{\bf
B}_{\mathbb{Q}}X_0,\prec_{EM})$. Furthermore, since the set
${\mathcal P}_{fin}{\bf B}_{\mathbb{Q}}X_0$ is countable, by
Proposition 2.2.3 in \cite{abramsky}, every round ideal in ${\mathcal
P}_{fin}{\bf B}_{\mathbb{Q}}X_0$ has a cofinal
$\prec_{EM}$-ascending subsequence and hence following the same
proof as in (1), one can prove that  $({\mathcal P}{\bf
B}_{\mathbb{Q}}X_0,\prec_{EM})$ is order-isomorphic to $({\bf
C}{\bf B}_{\mathbb{Q}}X_0,\prec_{EM})$. On the other hand, one
can  easily show that any $\prec_{EM}$-ascending subsequence
$(F_n)_{n>0}$ in ${\mathcal P}_{fin}{\bf B}X$ is
$\sim_{EM}$-equivalent to a $\prec_{EM}$-ascending subsequence
$(G_n)_{n>0}$ in ${\mathcal P}_{fin}{\bf B}_{\mathbb{Q}}X_0$ . Hence
$({\bf C}{\bf B}_{\mathbb{Q}}X_0,\prec_{EM})$ is order-isomorphic
to  $({\bf C}{\bf B}X,\prec_{EM})$. Therefore, the proof is
established.\qed

\begin{cor}
Let $(X,d)$ be a complete metric space. Then the Plotkin
powerdomain ${\mathcal P}{\bf B}X$ and $\omega$-Plotkin ${\bf C}{\bf
B}X$ are isomorphic.\qed
\end{cor}

\section{Future Work}

In this paper, various computational models of the hyperspace
${\mathcal K}_0(X)$ of the non-empty compact subsets of a
quasi-metric space $(X,d)$ were studied. It was shown how to use
a special computational model ${\bf B}X$ of $(X,d)$ to get the
corresponding $\omega$-computational model ${\bf C}{\bf B}X$ of
${\mathcal K}_0(X)$. The above construction would have been more
satisfactory if  a computational model for the ${\mathcal K}_0(X)$
could have been defined, starting from an arbitrary computational
model of $(X,d)$. This idea is already developed for the case of
metric spaces by Martin \cite{martin1}, by appealing to the
notion of a measurement, and for the spaces with countable based
models by Berger et. al. in \cite{berger}. Therefore, an
interesting subject of research is to find a fairly general
framework under which the ideas from \cite{martin1} and
\cite{berger} can be generalized to the present context. Another
topic of research is to study the effectiveness of ${\mathcal
K}_0(X)$. So one could ask whether ${\bf C}{\bf B}X$ supports an
effective base whenever $(X,d)$ is an effective quasi-metric.

Section 5 involves a generalization of the results obtained
earlier, showing that  ${\mathcal K}_0(X)$ has a quantitative
$\omega$-computational model. One would desire to have a
quantitative computational model for ${\mathcal K}_0(X)$. This, for
example, requires to generate another computational model from
$({\bf C}{\bf B}X,\sqsubseteq,D,\phi)$, similar to what we
obtained in Theorem \ref{domrep}. One way to achieve this is to
employ the Yoneda-completion of $({\bf C}{\bf B}X,D)$,  which
serves as a natural generalization of the ideal completion.

In section 6,  the Plotkin powerdomain and $\omega$-Plotkin
domain of ${\bf B}X$ were compared and the situations in which
both constructions are order-isomorphic were observed. Now, the
question of finding an example for which these constructions are
not isomorphic is imposed.

\section*{Acknowledgement} The authors would like to
thank R. Heckmann and O. Valero for their useful correspondence.
The authors would also like to thank two anonymous referees for
their constructive comments which enabled  us to correct some of the
results and improve the presentation of the paper.

\end{document}